\documentclass[11pt]{article}
\usepackage{framed,multirow}
\usepackage{wrapfig, booktabs, caption}
\usepackage{latexsym}
\usepackage{amsmath,amsthm,amsfonts,amssymb,mathrsfs,dsfont,breqn} 
\usepackage{autobreak}
\usepackage{multirow}
\usepackage{wrapfig, booktabs, caption}
\usepackage{comment}
\usepackage[dvipsnames]{xcolor}
\usepackage[margin=0.70in]{geometry}

\usepackage[
    backend=biber,
    style=authoryear,
    natbib=true,
    maxbibnames=3,
    minbibnames=1,
    maxcitenames=2,
    mincitenames=1,
    uniquelist=false,
    uniquename=false,
    giveninits=true,
    date=year,
]{biblatex}

\renewbibmacro{in:}{}

\DeclareNameAlias{default}{family-given}
\DeclareNameAlias{sortname}{family-given}
\DeclareNameAlias{author}{family-given}

\newcommand{\dtoprule}{\specialrule{1pt}{0pt}{1.0pt}
            \specialrule{0.3pt}{0pt}{\belowrulesep}
            }
\newcommand{\dbottomrule}{\specialrule{0.3pt}{1.5pt}{1.0pt}
            \specialrule{1pt}{0pt}{\belowrulesep}
            }
\newcommand{\quartquad}{\hspace{0.25em}} 
 
\usepackage{url}
\usepackage{xcolor}
\usepackage{soul}
\usepackage[citebordercolor=white]{hyperref}
\usepackage{fancyhdr}
\hypersetup{
    colorlinks=true,
    citecolor=blue,
    linkcolor=blue,
    urlcolor=blue,
}

\fancypagestyle{plain}{
    \fancyhf{}
    \fancyfoot[L]{%
    \footnotesize
    © 2026. This manuscript version is made available under the \href{https://creativecommons.org/licenses/by-nc-nd/4.0/}{CC-BY-NC-ND 4.0 license}.
    }
    \fancyfoot[R]{\thepage}
}

\title{
Analyzing Cosmic Ray Spectral Features:
A Numerical Investigation \thanks{
Accepted for publication on 21 August 2025 in the Special Issue "Astrophysics of Cosmic Rays in the Multi-messenger Era" of \textit{Advances in Space Research}.
The final published version is available at DOI:
\href{https://doi.org/10.1016/j.asr.2025.08.050}
{10.1016/10.1016/j.asr.2025.08.050}.
}
}

\author{
Yuca C. Chen$^{1,2}$ \thanks{Corresponding author: yucachen@udel.edu. Currently at the Bartol Research Institute, Department of Physics and Astronomy, University of Delaware, Newark, DE 19716, USA.}, 
Zachary M. Dorris$^{1,2}$,
Eun-Suk Seo$^{1,2}$ \thanks{Corresponding author: seo@umd.edu},
Vladimir S. Ptuskin$^{2}$\\[0.5em]
$^{1}${\small Department of Physics, University of Maryland, College Park, MD 20742, USA.}\\
$^{2}${\small Institute for Physical Science and Technology, University of Maryland, College Park, MD 20742, USA.}\\
}

\date{}

\begin{document}

\maketitle

\vspace{-8mm}

\begin{abstract}
Recent cosmic ray space-based and balloon-borne experiments have revealed various spectral features. Spectral hardening around $\sim$200 GV has been seen in primary nuclei as well as secondaries produced during propagation. Proton spectrum softening at $\sim$10 TV and helium spectrum softening at a few tens TV has also been seen. Additionally, a positron excess has been observed above $\sim$25 GeV. The cosmic ray propagation code, GALPROP v57, was utilized to investigate the cause behind these features. A diffusion model with reacceleration and convection effects was used as a baseline. To find the best fit to the experimental data, GALPROP v57’s parameter optimization module, utilizing the external numerical minimization software MINUIT2, was used. For the hardening, three scenarios were studied: (1) a diffusion coefficient break, (2) injection spectra breaks, and (3) a combination of both breaks. An additional injection spectrum break was considered to fit the softening of the proton and helium spectra. An additional positron source was introduced for the positron excess. The resulting elemental spectra and ratios, along with the all-particle spectrum, are compared to compiled cosmic ray data. Implications of these spectral features are also discussed.
\end{abstract}


\section{Introduction}

In recent times, high-precision cosmic ray space-based and balloon-borne experiments have revealed spectral features in different cosmic ray species \citep[][\& references therein]{2021JKPS...78..923S}. 
These features include a spectral hardening at $\sim$200 GV, observed by ATIC-2 \citep{Panov_2009} and reported by CREAM \citep{2010ApJ...714L..89A, 2011ApJ...728..122Y}, PAMELA \citep{2011Sci...332...69A}, AMS-02 \citep{PhysRevLett.114.171103, PhysRevLett.119.251101, PhysRevLett.120.021101, PhysRevLett.121.051103, PhysRevLett.124.211102, PhysRevLett.126.081102, PhysRevLett.126.041104, PhysRevLett.130.211002}, DAMPE \citep{2019SciA....5.3793A, 2021PhRvL.126t1102A}, and CALET \citep{PhysRevLett.125.251102, PhysRevLett.126.241101, PhysRevLett.129.101102, PhysRevLett.130.171002}.
At higher energies, a softening in the proton data at $\sim$10 TeV was first reported by CREAM \citep{2017ApJ...839....5Y} and then by DAMPE \citep{2019SciA....5.3793A}, ISS-CREAM \citep{2022ApJ...940..107C}, and CALET \citep{PhysRevLett.130.171002}. A softening in the helium data has also been reported at slightly higher energies compared to the proton data, also initially reported by CREAM \citep{2017ApJ...839....5Y} and then by DAMPE \citep{2021PhRvL.126t1102A} and CALET \citep{PhysRevLett.130.171002}. Excesses have been observed in the electron data, with ATIC first reporting an excess in the all-electron data at energies of $\sim$300–800 GeV \citep{2008Natur.456..362C}. PAMELA \citep{2009Natur.458..607A} and AMS-02 \citep{PhysRevLett.122.041102} later reported an excess in the positron data above $\sim$25 GeV. 

Studying these spectral features can provide insight into the mechanisms that govern the acceleration and propagation of these species. In this work, we use the cosmic ray propagation code GALPROP \citep{Strong_1998, Moskalenko_1998} to study these features by fitting cosmic ray data. Three different scenarios are tested for the spectral hardening: (1) a diffusion coefficient break, (2) injection spectra breaks, and (3) a combination of both breaks. An additional injection spectrum break is considered for the softening in the proton and helium data. A charge-symmetric positron primary source is considered for the positron excess. The results are then compared to compiled cosmic ray spectra and ratio data.  After fitting, the results from GALPROP for all cosmic ray species are summed to obtain an all-particle spectrum. The result is then compared to compiled all-particle spectrum data.

This paper is organized as follows: Section \ref{modelSect} describes the GALPROP cosmic ray propagation code, Section \ref{methoSect} describes the methodology for fitting data, Section \ref{ResultsSection} describes the results, Section \ref{discussSect} provides discussion of the results, and Section \ref{concluSect} provides concluding remarks.

\section{Model} \label{modelSect}

\subsection{GALPROP Cosmic Ray Propagation Code}

For this work, the cosmic ray propagation code GALPROP v57 is utilized to fit cosmic ray data \citep{Porter_2022}. With a given source distribution and boundary conditions, GALPROP v57 simultaneously solves partial differential equations that describe particle transport for many cosmic ray species. The equation describing this transport is given in Equation \ref{CRPropEquation} \citep{Strong_1998}. 
On the right-hand side, going from left to right, the equation includes terms for source distribution, diffusion with a spatial diffusion coefficient, diffusive reacceleration, energy loss, and fragmentation and radioactive decay effects. Energy loss effects due to ionization and Coulomb scattering, inverse Compton scattering, bremsstrahlung radiation, and synchrotron radiation are included.  Secondary particle and isotope production is also included, alongside electron pickup, electron K-capture, knock-on electrons, and stripping processes. Additional details and processes concerning GALPROP v57 can be found in several publications \citep[\& references therein]{Porter_2022}.

\vspace{-2.70mm}

\begin{multline}
    \frac{\partial \psi}{\partial t} = q(\Vec{r},p) + \Vec{\nabla} \cdot (D_{xx} \Vec{\nabla} \psi - \Vec{V} \psi) + \frac{\partial}{\partial p} p^{2} D_{pp} \frac{\partial}{\partial p} \frac{1}{p^{2}} \psi
    - \frac{\partial}{\partial p} \left[ \dot p \psi - \frac{p}{3} (\Vec{\nabla} \cdot \Vec{V}) \psi \right] - \frac{1}{\tau _{f}} \psi - \frac{1}{\tau _{r}} \psi 
\label{CRPropEquation}
\end{multline}

The spatial diffusion coefficient is proportional to rigidity $\rho$ by $D_{xx} \propto \beta D_{0} \rho^{ \delta}$, where $\beta = v \slash c$, $D_{0}$ is the diffusion coefficient value at a normalization rigidity, and $\delta$ is the diffusion coefficient index. If a break in the spatial diffusion coefficient is included, the coefficient is taken as $D_{xx} \propto \beta D_{0} (\rho \slash \rho_{0})^{\delta_{0}}$ with index $\delta_{0}$ below the break rigidity $\rho_0$ and $D_{xx} \propto \beta D_{0} (\rho \slash \rho_{0})^{\delta_{1}}$ with index $\delta_{1}$ above the break rigidity, with $\delta_{0} \neq \delta_{1}$. When “minimal” reacceleration is involved (i.e. only accounting for stochastic reacceleration of cosmic rays by interstellar turbulence), the diffusion coefficient in momentum space $D_{pp}$ is related to the spatial diffusion coefficient $D_{xx}$ through the Alfv\'en velocity $v_{Alfv\acute{e}n}$  \citep{1990acr..book.....B, 1994ApJ...431..705S}. 

In GALPROP v57, a single break in the spatial diffusion coefficient is included. However, in this work, a second spatial diffusion coefficient break is incorporated to explore the possibility of having two breaks. An additional break in rigidity $\rho_1$, with $\rho_0 < \rho_1$, is implemented. An accompanying diffusion coefficient index $\delta_2$ is also included, where $\delta_2$ is the index where $\rho >\rho_1$. $\delta_1$ now denotes the index when $\rho_0 < \rho <\rho_1$. The term $\zeta$ in $D_{pp}$ when minimal reacceleration is involved, shown in Equation \ref{DppRelation} \citep{1994ApJ...431..705S}, is adjusted accordingly from these inclusions. $\zeta$ is defined as the logarithmically weighted average of the spatial diffusion coefficient indices between 1 MV and 1 PV, and it is adjusted to the expression in Equation \ref{deltaConstantEquation} for when 1 MV $ < \rho < $ 1 PV.

\begin{equation}
    D_{pp} \propto \frac{4}{3 \zeta(4 - \zeta)(4 - \zeta^{2})}
\label{DppRelation}
\end{equation}

\begin{equation}
    \zeta = \frac{\delta_0 \log{\frac{\rho_0}{1 \quartquad \mathrm{MV}}} + \delta_1 \log{\frac{\rho_1}{\rho_0}} + \delta_2 \log{\frac{1 \quartquad \mathrm{PV}}{\rho_1}}}{\log{(1 \quartquad \mathrm{PV}})}
\label{deltaConstantEquation}
\end{equation}

For a source distribution, the injection spectra of cosmic ray species follow a power-law in rigidity $R$ by $q(R) \propto (R \slash R_i)^{- \gamma_j}$. $R_i$ are the breaks in rigidity and $\gamma_j$ are the spectral indices, where $i =0, 1, 2$ and $j = 0, 1, 2, 3$. The spectral index $\gamma_0$ applies for $R<R_0$, $\gamma_1$ for $R_0 < R < R_1$, $\gamma_2$ for $R_1 < R < R_2$, and $\gamma_3$ for $R > R_2$. 

The proton and negative electron fluxes are independently normalized at distinct kinetic energies. In this work, the proton flux is normalized at 100 GeV and the negative electron flux is normalized at 70 GeV, both based on available AMS-02 data \citep{AGUILAR20211}. The remaining cosmic ray nuclei and $\bar p$ are then produced using relative elemental and isotopic abundances with respect to protons. The specific relative abundances used for this work are reported in Section \ref{ResultsSection} in Results.

\subsection{MINUIT2 Minimization Library}

The diffusion coefficient and injection spectra results are determined by fitting cosmic ray data. To do this, GALPROP v57’s parameter optimization module is used, which utilizes the minimization library MINUIT2 \citep{Porter_2022}.  
MINUIT2 returns best-fit parameter values, along with their uncertainties, once the minimum value of the input multi-parameter objective function is found \citep{JAMES1975343}. To optimize the parameter values, experimental data must be input into MINUIT2. In this work, the MIGRAD algorithm is used to minimize the chi-square loss between the output from GALPROP v57 and the input experimental data. Additional details on the MIGRAD algorithm can be found in \citet{JAMES1975343}.

\section{Methodology} \label{methoSect}

For this study, a diffusion model with reacceleration and convection effects was considered. Parameter values from the example \textit{galdef} file "galdef\_57\_CRfit\_SNR" provided in GALPROP v57 were used as a base set. The example supernova remnant spatial profile "snr\_src.xml", also provided in GALPROP v57, was used with the \textit{galdef} file as part of the base set of parameter values. In addition to tuning the free parameters in the base model, the solar modulation potential $\phi$ from the force field approximation was also considered as another free parameter \citep{1968ApJ...154.1011G}. The cross-section option “Opt 022” in GALPROP v57 was used during the fitting procedure as well as a static halo half-width of 7.5 kpc.

Simultaneously fitting all the data across its entire rigidity range caused the optimizer to return a poor fit to the data, despite minimizing the chi-squared loss. 
Data fitting was then divided into four sequential stages to encourage the optimizer to return a better fit. Stage 1 involved building up a base model of spectra up from $\sim$1 MV up to $\rho_1$ or $R_1$. Stage 2 included higher rigidity data to fit data up to $R_2$. Stage 3 included higher rigidity p and He data above $R_2$ to fit these data based on the results from Stage 2. Stage 4 examined the positron data also based on results from Stage 2. For all of these stages, data were input in the USINE format \citep{USINEFormat} as required by GALPROP v57’s parameter optimization routine \citep{Porter_2022}. 

Our prior work used GALPROP v56 to fit the spectral hardening for elements p through O, the p and He softening, and the positron excess with a primary source through hand-tuning parameter values \citep{Chen:2023eyy}. In this paper, we aim to fit additional heavy elemental spectra data of Ne through Fe, introduce a charge-symmetric primary positron source if necessary, calculate the resulting all-particle spectrum, and obtain errors for propagation parameters related to spectral features using GALPROP v57's parameter optimization routine.

\subsection{Fitting Data from $\sim$1 MV up to $\rho_1$ or $R_1$} \label{Stage1MethodSect}

For Stage 1, a base set of diffusion coefficient values were first determined up to $\rho_1$. For this, the data of B/C, B/O, B, C, O, negative $e^{-}$, and $e^{+}$ from the AMS-02 experiment \citep{AGUILAR20211}, and their respective available local interstellar data from the Voyager1 experiment \citep{2013Sci...341..150S, 2016ApJ...831...18C} were first input. The B/C, B, and C data from the ACE-CRIS experiment \citep{2006AdSpR..38.1558D} were also input for developing the base diffusion coefficient values. These input AMS-02 data were taken from the 2011/05-2018/05 collection period and were selected due to their statistical significance. The input ACE-CRIS data were taken from the 1998/01-1999/01 collection period and were selected for additional diffusion coefficient information. The input Voyager1 data were selected for local interstellar spectra information. The input data for the elemental spectra and ratios were limited to be below $\sim$200 GV for the optimizer to focus on this rigidity range. The negative $e^{-}$ data was limited to be below $\sim$42 GeV \citep{PhysRevLett.122.101101}, and the $e^{+}$ data was limited to be below $\sim$25 GeV \citep{PhysRevLett.122.041102}.

After obtaining a baseline set of diffusion coefficient values, the elements of p, $\bar{p}$, He, Be, C, N, O, Ne, Mg, Si, S, and Fe from the AMS-02 experiment \citep{PhysRevLett.124.211102, PhysRevLett.126.041104, AGUILAR20211, PhysRevLett.130.211002} and their respective available interstellar data from the Voyager1 experiment \citep{2013Sci...341..150S, 2016ApJ...831...18C, 2019NatAs...3.1013S} were input. 
All species were initially put into a single injection spectrum class up to $R_1$. The elements were then split up into different injection spectra classes when deemed necessary. Splitting the elements into different classes followed a strategy based on Section 5.4 of "Fixing and Releasing Parameters" from the MINUIT handbook \citep{James:1994vla}. In this strategy, previously established parameters from convergent optimization runs would first be “fixed” (i.e. not allowed to vary as a free parameter). New free parameters would then be introduced as free parameters for the optimizer to vary. Another optimization run would then follow, where these previously fixed parameters and newly introduced varying free parameters were declared together, denoted as a "semi-fixed" run. Once the semi-fixed run converged, values for these free parameters would then be obtained, denoted as “semi-optimized” values. Afterwards, all parameters were then reintroduced as free parameters to the optimizer with either their previously established or semi-optimized values. All the parameters would then be allowed to vary simultaneously to obtain a valid set of values and associated errors. This strategy was then repeated in a loop until all elements were fit below $R_1$.

\subsection{Fitting Data up to $R_2$} \label{Stage2MethodSect}

To extend the fit range up to $R_2$, the diffusion coefficient and injection spectra results from Stage 1 were fixed. Data from the AMS-02 \citep{PhysRevLett.124.211102, PhysRevLett.126.041104, AGUILAR20211,  PhysRevLett.130.211002}, CALET \citep{PhysRevLett.129.251103, PhysRevLett.129.101102, PhysRevLett.130.171002}, CREAM-I \citep{2008APh....30..133A}, CREAM-I+III \citep{2017ApJ...839....5Y}, CREAM-II \citep{2009ApJ...707..593A}, DAMPE \citep{2019SciA....5.3793A, 2021PhRvL.126t1102A, DAMPECOLLABORATION20222162}, ISS-CREAM \citep{2022ApJ...940..107C}, and NUCLEON-KLEM \citep{2019AdSpR..64.2546G} experiments were input. 
CREAM-II and NUCLEON-KLEM data of N, Ne, Mg, Si, and Fe along with CALET data of B, C, O, and Fe were not used due to absolute normalization issues between the AMS-02 data \citep{PhysRevLett.125.251102, PhysRevLett.126.241101, PhysRevLett.129.251103, PhysRevLett.124.211102, PhysRevLett.126.041104}.
Table \ref{Stage2InputData} summarizes the data used for Stage 2 along with their input rigidity range.

As in \citet{Wu:2021ldb}, three cases were tested in order to fit data up to $R_2$: one with a diffusion coefficient break (Case 1), one with injection spectra breaks (Case 2), and one with a combination of both effects (Case 3). The diffusion coefficient break $\rho_1$, injection spectra break $R_1$, and corresponding indices above these breaks $\delta_2$ and $\gamma_2$ were allowed to be free parameters during fitting. In contrast to \citet{Wu:2021ldb}, a parameter optimization routine is used to obtain results for these three cases, along with using more recent data, including He data from CALET and S data from AMS-02. Additional differences between this study and \citet{Wu:2021ldb} include using the parameter optimization module to fit the p and He softening and a charge-symmetric primary $e^{+}$ source if deemed necessary and a comparison to the all-particle spectrum.

\subsection{Extending the Fit of the p and He Data Above $R_2$} \label{methoPandHeSection}

After Stage 2 was completed, the p and He fit was extended above $R_2$. This was done as the p and He data extends to higher rigidities in comparison to other species. 
The p data from the CALET \citep{PhysRevLett.129.101102}, CREAM-I+III \citep{2017ApJ...839....5Y}, DAMPE \citep{2019SciA....5.3793A}, ISS-CREAM \citep{2022ApJ...940..107C}, and NUCLEON-KLEM \citep{2019AdSpR..64.2546G} experiments were input. 
The He data from the CALET \citep{PhysRevLett.130.171002}, CREAM-I+III \citep{2017ApJ...839....5Y}, DAMPE \citep{2021PhRvL.126t1102A}, and NUCLEON-KLEM \citep{2019AdSpR..64.2546G} experiments were also input.
The input data were restricted to be above $\sim$200 GV. 
An additional injection spectrum break for p and He was used to fit these data.

\begin{table}[t]
    \centering
    \caption{Input data used for Stage 2 of the Methodology.}
    \begin{tabular}{ccc} \dtoprule
         Species & Experiment & Input Range\\ \midrule
         p, $\bar{p}$, He, &  &  \\
         Be, B, B/C, B/O, &  & Entire  \\
         C, N, O, &  AMS-02 &  Rigidity \\ 
         Ne, Mg, Si, S, Fe &  & Range \\
         Neg. $e^{-}$, and $e^{+}$ &  & \\ \midrule
         p$^*$, He$^*$, and B/C & CALET &  \\ 
         B/C and B/O & CREAM-I & \\
         p$^*$ and He$^*$ & CREAM-I+III &  \\ 
         C and O & CREAM-II & $\geq$ 200 GV \\ 
         p$^*$, He$^*$, B/C, and B/O & DAMPE & \\ 
         p$^*$ & ISS-CREAM & \\  
         p$^*$, He$^*$, C$^*$, and O$^*$ & NUCLEON-KLEM &  \\ \dbottomrule
         \multicolumn{3}{l}{\footnotesize $^*$ Data was input up to reported softening.}
    \end{tabular}
    \label{Stage2InputData}
    \vspace{1.5mm}
\end{table}

\subsection{Extending the Fit of the Positron Data} \label{methoPosSection}
Following Stage 2, the fit of the positron data was extended by introducing a charge-symmetric primary positron source to fit the data. The example supernova remnant spatial profile "snr\_src.xml" provided in GALPROP v57 was used as a base for the charge-symmetric primary positron source.
The data of $e^{-}$ and $e^{+}$ from the AMS-02 experiment \citep{AGUILAR20211} were input.

\section{Results} \label{ResultsSection}

Using GALPROP v57’s parameter optimization module, the best-fit parameters and their associated error for each case are obtained from $\sim$1 MV up to $\rho_1$ and $R_1$ for the elements of p, $\bar p$, He, Be, B, C, B/C, N, O, Ne, Mg, Si, S, Fe, negative $e^-$, and $e^+$. These results are the outcome of Stage 1 from Section \ref{Stage1MethodSect}.

For all three cases, the resulting diffusion coefficient is $D_{xx} \propto \beta D_{0} (\rho \slash \rho_{0})^{\delta_0}$ below $\rho_{0}$, where $D_{0} = (7.81 \pm 0.04)$ $\cdot $10$^{28}$ cm$^{2}$ s$^{-1}$ at $R = 10$ GV, $\rho_{0} = 6.78 \pm 0.06$ GV, and $\delta_{0} = 0.101 \pm 0.005$. Above $\rho_{0}$ and up to $\rho_{1}$, the diffusion coefficient is $D_{xx} \propto \beta D_{0} (\rho \slash \rho_{0})^{\delta_1}$, where $\delta_{1} = 0.493 \pm 0.001$ with the same values of $D_{0}$ and $\rho_{0}$. 
An Alfv\'en velocity $v_{Alfv\acute{e}n}$ of $16.9 \pm 0.2$ km s$^{-1}$ and a gradient of convection velocity $dV_{conv}/dz$ is $3.9 \pm 0.2$ is obtained. These results are summarized in Table \ref{diffCoeffParam}. Two solar modulation parameters are also obtained: $\phi_{AMS}$ = $533 \pm 2$ MV for the AMS-02 data and $\phi_{ACE}$ = $500 \pm 10$ MV for the ACE-CRIS data.

For all three cases below $R_1$, the injection spectral break rigidity $R_{0}$, the spectral index $\gamma_{0}$ below $R_{0}$, and the spectral index $\gamma_{1}$ above $R_{0}$ 
for each element and negative $e^{-}$ are summarized in Table \ref{baseInjectionSpectra}. The results in this table are applied to every isotope for each respective species. The same injection spectra results are found for C, N, and O, where these three elements are grouped together with the same injection spectral indices $\gamma_0$ and $\gamma_1$ and break rigidity $R_0$. The same injection spectra results are found for Ne, Si, and S as well. Above $R_0$, all injection spectral indices $\gamma_{1}$ are $\sim$2.3 except for $e^{-}$: $e^{-}$ is $\sim$0.3 softer with $\gamma_1$ at $2.655 \pm 0.001$. The break $R_{0}$ generally increases as the elements get heavier. Table \ref{abundancesTable} summarizes the primary source abundances in relation to the abundance for p. These abundances are kept the same for all later stages of fitting.

\begin{table}[h]
    \centering
    \caption{Diffusion coefficient results for all three cases.}
    \begin{tabular}{cc} \dtoprule
         $D_{0}$ (10$^{28}$ cm$^{2}$ s$^{-1}$) $(R = 10 \, \text{GV})$ & $7.81 \pm 0.04$\\
         $\delta_{0}$ & $0.101 \pm 0.005$ \\
         $\rho_{0}$ (GV) & $6.78 \pm 0.06$ \\
         $\delta_{1}$ & $0.493 \pm 0.001$ \\ 
         $v_{Alfv\acute{e}n}$ (km s$^{-1}$) & $16.9 \pm 0.2$ \\
         $dV_{conv}/dz$ (km s$^{-1}$ kpc$^{-1}$) & $3.9 \pm 0.2$\\ \dbottomrule
    \end{tabular}
    \label{diffCoeffParam}
\end{table}

\begin{table}[h]
    \centering
    \caption{Injection spectra results for all three cases.}
    \begin{tabular}{cccc} \dtoprule
         & $\gamma_{0}$ & $R_{0}$ (GV) & $\gamma_{1}$ \\ \midrule  
         p & $1.690 \pm 0.004$ & $1.93 \pm 0.03$ & $2.3417 \pm 0.0002$ \\
         He & $1.565 \pm 0.009$ & $2.44 \pm 0.03$ & $2.280 \pm 0.001$ \\
         C, N, O & $1.137 \pm 0.022$ & $2.01 \pm 0.03$ & $2.323 \pm 0.001$ \\
         Ne, Si, S & $1.570 \pm 0.023$ & $3.87 \pm 0.10$ & $2.322 \pm 0.002$ \\
         Mg & $1.463 \pm 0.041$ & $3.15 \pm 0.15$ & $2.355 \pm 0.003$ \\
         Fe & $1.986 \pm 0.014$ & $11.4 \pm 0.5$ & $2.311 \pm 0.006$ \\
         Neg. $e^{-}$ & $1.474 \pm 0.015$ & $4.21 \pm 0.03$ & $2.655 \pm 0.001$ \\ \dbottomrule
    \end{tabular}
    \label{baseInjectionSpectra}
\end{table}

\begin{table}[!h]
    \centering
    \caption{Primary source abundance results for all three cases.}
    \begin{tabular}{cc | cc} \dtoprule
         Isotope & Abundance & Isotope & Abundance \\ \toprule
         \textsuperscript{1}H & 1.00 $\cdot$ 10\textsuperscript{6} & \textsuperscript{4}He & $101321 \pm 172$ \\  \midrule
         \textsuperscript{12}C & $3272 \pm 9$ & \textsuperscript{14}N & $288 \pm 4$ \\  \midrule
         \textsuperscript{16}O & $4171 \pm 12$ & \textsuperscript{20}Ne & $451 \pm 4$ \\  \midrule
         \textsuperscript{24}Mg & $595 \pm 6$ & \textsuperscript{28}Si & $728 \pm 4$ \\  \midrule
         \textsuperscript{32}S & $113 \pm 1$ &\textsuperscript{56}Fe & $669 \pm 9$ \\  \dbottomrule
    \end{tabular}
    \label{abundancesTable}
    \vspace{1.5mm}
\end{table}

The results of Stage 2 from Section \ref{Stage2MethodSect} are split into three cases. For Case 1,  
the diffusion coefficient is found to have a break at $\rho_{1}$ = $226 \pm 9$ GV, with an index of $\delta_{1} = 0.493 \pm 0.001$ below and $\delta_{2}$ = $0.382 \pm 0.003$ above $\rho_{1}$. The diffusion coefficient index is smaller by $\sim$0.11 above $\rho_1$.
No injection spectra breaks are included for Case 1, so the injection spectral indices remain as $\gamma_1$ in Table \ref{baseInjectionSpectra} above the break $\rho_1$. 

In comparison to Case 1, Case 2 includes only injection spectra breaks to fit the data.
The injection spectral break rigidity $R_1$ and spectral index $\gamma_2$ above $R_1$ for each element and negative $e^-$ are summarized in Table \ref{hardeningSourceInjection}. The elements of He, C, N, O, and Fe are found to have the same $R_1$ at $478 \pm 23$ GV and $\gamma_2$ of $2.059 \pm 0.010$, comprising an element group. The elements of Ne, Mg, Si, and S are found to have the same $R_1$ at $463 \pm 95$ GV and $\gamma_2$ of $2.157 \pm 0.064$, comprising another element group. $R_1$ for these two element groups is found to be consistent within errors. All indices $\gamma_2$ are smaller above the break point $R_1$ compared to $\gamma_1$ below: spectra become harder by $\sim$0.2 above $R_1$. 
The injection spectrum for p has a break $R_1$ at $338 \pm 20$ GV, which is found to be $\sim$125-150 GV lower compared to the heavier elements. From this, the spectrum for p hardens at a lower rigidity compared to the heavier elements. The elements of He, C, N, O, and Fe are found to have a smaller $\gamma_2$ compared to the other elements of p, Ne, Mg, Si, and S by $\sim$0.1: the spectra of He, C, N, O, and Fe harden more compared to the other elements. The break $R_1$ for $e^-$ is $\sim$170-310 GV lower compared to other elements, and the index $\gamma_2$ for $e^-$ is larger by $\sim$0.25 compared to other elements. The spectrum for $e^-$ hardens less and at a lower rigidity compared to other elements. No diffusion coefficient break is included for Case 2, so the diffusion coefficient index remains as $\delta_1$ in Table \ref{diffCoeffParam} above $R_1$.

Case 3 includes both a high-rigidity diffusion coefficient break and injection spectra breaks to fit the data. 
The injection spectral break rigidity $R_1$ and spectral index $\gamma_2$ above $R_1$ for each element and negative $e^-$ are summarized and compared in Table \ref{hardeningSourceInjection} for Cases 2 and 3. All indices $\gamma_2$ in Case 3 are smaller above the break point $R_1$ compared to $\gamma_1$ below: spectra become harder by $\sim$0.2 above $R_1$. 
The same injection spectra element groups from Case 2 are found for Case 3. These groups are the same as those found by AMS-02 for the spectral hardening \citep{PhysRevLett.119.251101, PhysRevLett.124.211102, PhysRevLett.126.041104, PhysRevLett.130.211002}. 
The injection spectrum for p has a break $R_1$ at $553 \pm 30$ GV for Case 3, which is found to be $\sim$130-160 GV lower in comparison to the heavier elements. The spectrum for p hardens at a lower rigidity compared to the heavier elements, as also obtained in Case 2. The break $R_1$ for $e^-$ is $\sim$400-560 GV lower compared to other elements, and the index $\gamma_2$ for $e^-$ is larger by $\sim$0.30 compared to other elements. The spectrum for $e^-$ hardens less and at a lower rigidity compared to other elements, as also obtained in Case 2. For all nuclei, $R_{1}$ for Case 3 is found to have a higher rigidity compared to Case 2: spectra for Case 3 harden $\sim$200-250 GV above the break $R_{1}$ for Case 2.  
The spectral index $\gamma_2$ for p, He, C, N, O, and Fe in Case 3 are larger by $\sim$0.02 compared to Case 2: these spectra harden less in Case 3 compared to Case 2. However, the index $\gamma_2$ for Ne, Mg, Si, and S in Cases 2 and 3 is found to be consistent with each other within errors. Additionally, the break $R_1$ and index $\gamma_2$ for $e^-$ in Cases 2 and 3 are found to be consistent with each other within errors. Including a high-rigidity diffusion coefficient break, in addition to injection spectra breaks, has minimal impact on the amount of hardening in the injection spectrum of Ne, Mg, Si, and S. It also has minimal impact on the injection spectrum for $e^-$.

\begin{table}[h] 
    \centering
    \caption{Injection spectra results for Cases 2 and 3.}
    \begin{tabular}{cccc} \dtoprule
         &  &  Case 2 & Case 3\\ \midrule
         p & $R_{1}$ (GV) &  $338 \pm 20$ & $553 \pm 30$ \\
         & $\gamma_{2}$ &  $2.163 \pm 0.007$ & $2.186 \pm 0.008$ \\ \midrule
         He, C, N,& $R_{1}$ (GV) &  $478 \pm 23$ & $687 \pm 49$ \\
         O, Fe& $\gamma_{2}$ &  $2.059 \pm 0.010$ & $2.087 \pm 0.009$ \\ \midrule
         Ne, Mg, & $R_{1}$ (GV) &  $463 \pm 95$ & $714 \pm 99$ \\
         Si, S& $\gamma_{2}$ &  $2.157 \pm 0.064$ & $2.148 \pm 0.067$ \\ \midrule
         Neg. $e^{-} $& $R_{2}$ (GV) &  $169 \pm 17$ & $156 \pm 6$ \\
         & $\gamma_{2}$ &  $2.378 \pm 0.035$ & $2.405 \pm 0.012$ \\ \dbottomrule
    \end{tabular}
    \label{hardeningSourceInjection}
    \vspace{1.5mm}
\end{table}

The diffusion coefficient break $\rho_{1}$ and the index $\delta_{2}$ above $\rho_{1}$ for Cases 1 and 3 are compared in Table \ref{hardeningDiffusionCoeff}.
The diffusion coefficient has a break at $\rho_{1}$ = $201.8 \pm 0.3$ GV, with an index of $\delta_{1} = 0.493 \pm 0.001$ below and $\delta_{2}$ = $0.446 \pm 0.003$ above $\rho_{1}$. 
The diffusion coefficient index is smaller by $\sim$0.05 above $\rho_1$, reduced by $\sim$0.06 from Case 1. 
The break rigidity $\rho_{1}$ in Case 3 is $\sim$25 GV lower than in Case 1.

\begin{table}[h]
    \centering
    \caption{Diffusion coefficient results for Cases 1 and 3.}
    \begin{tabular}{ccc} \dtoprule
         &  Case 1 & Case 3\\ \midrule
         $\rho_{1}$ (GV) & $226 \pm 9$ & $201.8 \pm 0.3$ \\  \midrule
         $\delta_{2}$ & $0.382 \pm 0.003$ & $0.446 \pm 0.004$ \\  \dbottomrule
    \end{tabular}
    \label{hardeningDiffusionCoeff}
    \vspace{1.5mm}
\end{table}

For each element, the break $R_1$ for Case 2 in Table \ref{hardeningSourceInjection} is in between $\rho_1$ and $R_1$ for Case 3 due to a gradual hardening in the data as reported by AMS-02 \citep{PhysRevLett.114.171103, PhysRevLett.119.251101, PhysRevLett.120.021101, PhysRevLett.121.051103, PhysRevLett.124.211102, PhysRevLett.126.041104, PhysRevLett.130.211002}. No smoothing parameters were used for the diffusion coefficient and injection spectra power law in this work, which serves as a potential limitation in this study. 

\begin{figure}[!t]
    \centering
    \includegraphics[width=0.56\textwidth]{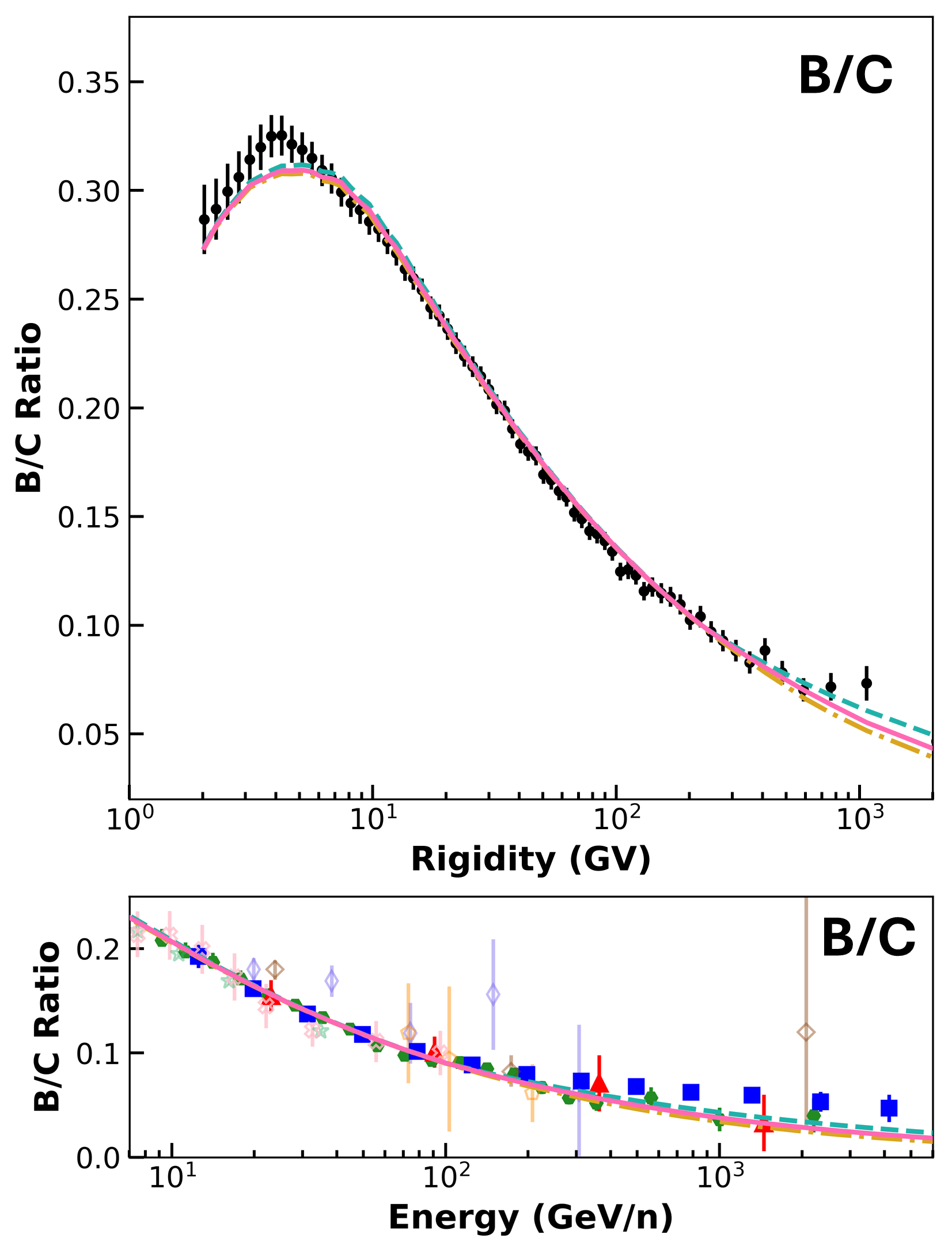}
    \caption{The results for the B/C ratio data for Cases 1 (dashed teal line), 2 (dashed-dotted gold line), and 3 (solid pink line) are compared with the compiled data. 
    The results are modulated spectra with $\phi_{AMS}$ = $533 \pm 2$ MV. 
    Direct measurement data legend: black filled dots -- AMS-02 \citep{AGUILAR20211}, periwinkle open thin diamonds -- ATIC-2 \citep{2008ICRC....2....3P}, green filled hexagons -- CALET \citep{PhysRevLett.129.251103}, red filled triangles -- CREAM I \citep{2008APh....30..133A}, beige open pentagons -- CRN \citep{1991ApJ...374..356M}, blue filled squares -- DAMPE \citep{DAMPECOLLABORATION20222162}, light green open stars -- HEAO3-C2 \citep{1990A&A...233...96E}, pink open wide crosses -- PAMELA \citep{2014ApJ...791...93A}, and brown open wide diamonds -- TRACER \citep{2011ApJ...742...14O}. ATIC-2, CRN, HEAO3-C2, PAMELA, and TRACER data are shown for additional comparison along with the data used for optimization.}
    \label{BCRatioFig}
\end{figure}

\begin{figure}[!h]
    \centering
    \includegraphics[width=0.56\textwidth]{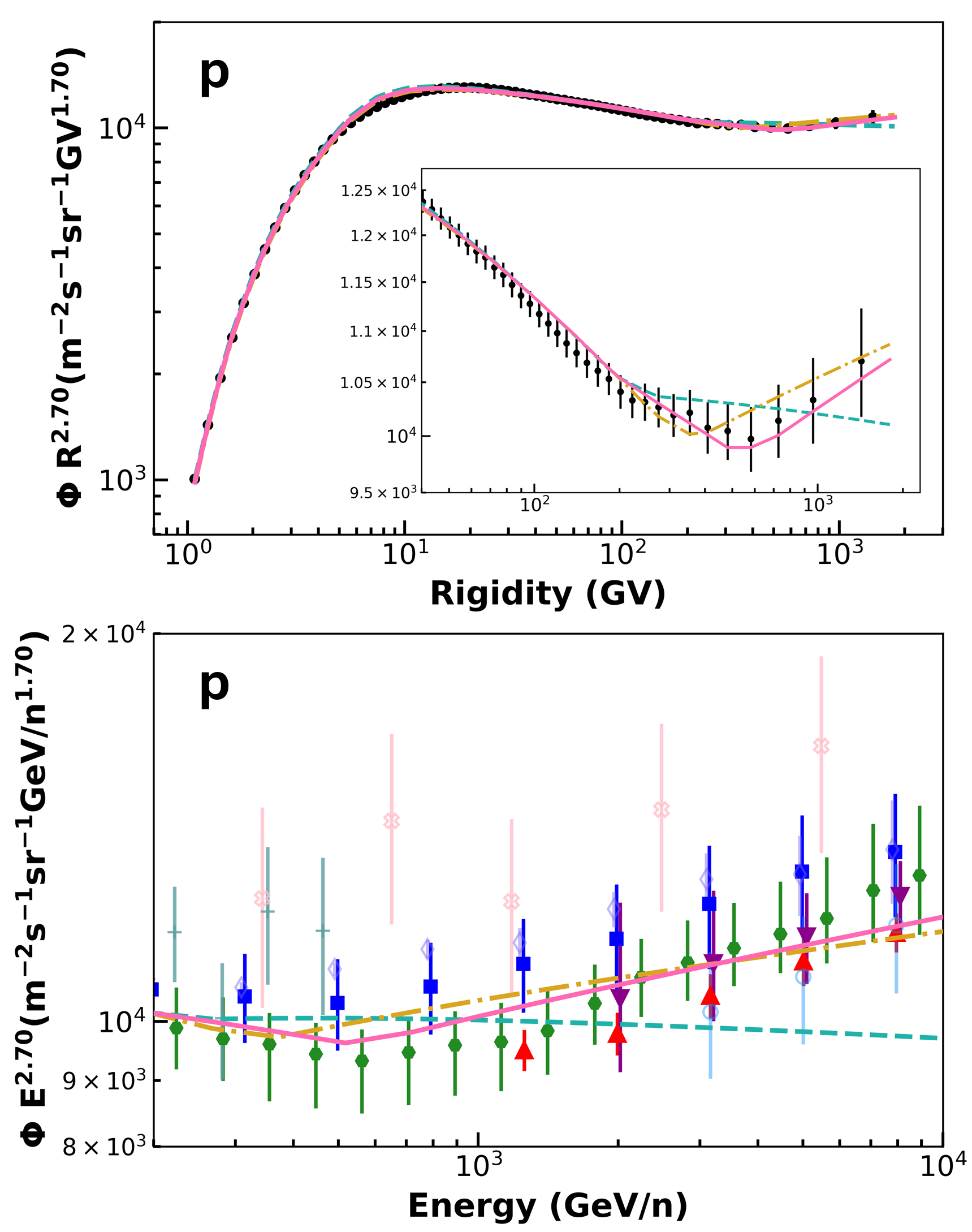}
    \caption{ The results for the p data for Cases 1 (dashed teal line), 2 (dashed-dotted gold line), and 3 (solid pink line) are compared with the compiled data.  
    A zoomed-in graph is shown above 40 GV in the top graph, also compared to the compiled data. The results are modulated spectra with $\phi_{AMS}$ = $533 \pm 2$ MV. 
    The legend can be found in Figure \ref{BCRatioFig} with additional entries: teal thin pluses -- BESS \citep{2007APh....28..154S}, purple downward-pointing filled triangles -- ISS-CREAM \citep{2022ApJ...940..107C}, and light blue open circles -- NUCLEON-KLEM \citep{2019AdSpR..64.2546G}. 
    The red upward-pointing filled triangles denote CREAM I+III \citep{2017ApJ...839....5Y} data in this figure. BESS data, along with ATIC-2 and PAMELA data, are shown for additional comparison along with the data used for optimization.
    }
    \label{pHardeningFig}

    \includegraphics[width=0.54\textwidth]{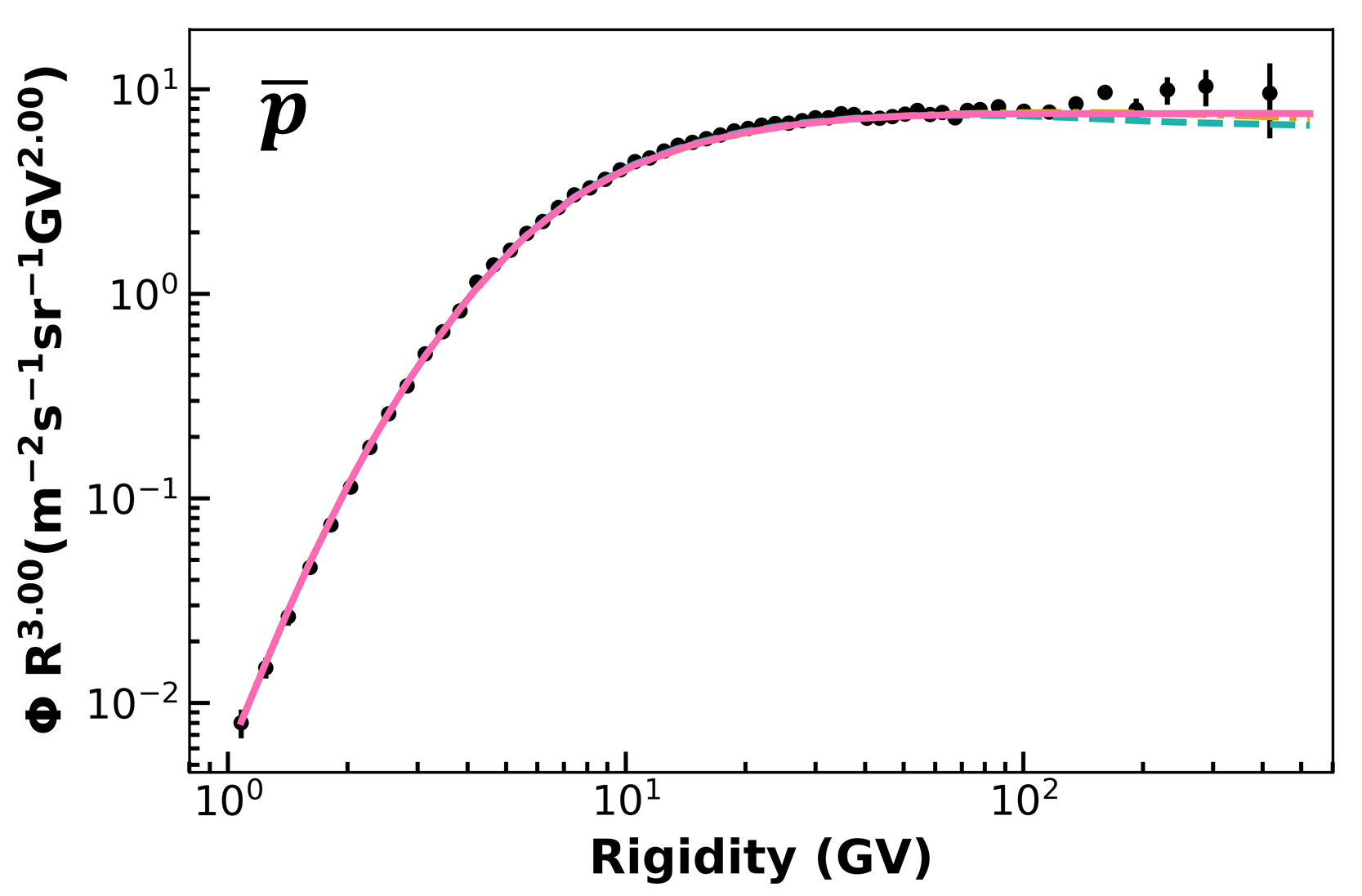}
    \caption{ The results for the $\bar p$ data for Cases 1 (dashed teal line), 2 (dashed-dotted gold line), and 3 (solid pink line) are compared with the compiled data, the legend for which can be found in Figure \ref{BCRatioFig}.
    The results are modulated spectra with $\phi_{AMS}$ = $533 \pm 2$ MV.
    }
    \label{pbarFig}
\end{figure}

\begin{figure}[!h]
    \centering
    \includegraphics[width=0.56\textwidth]{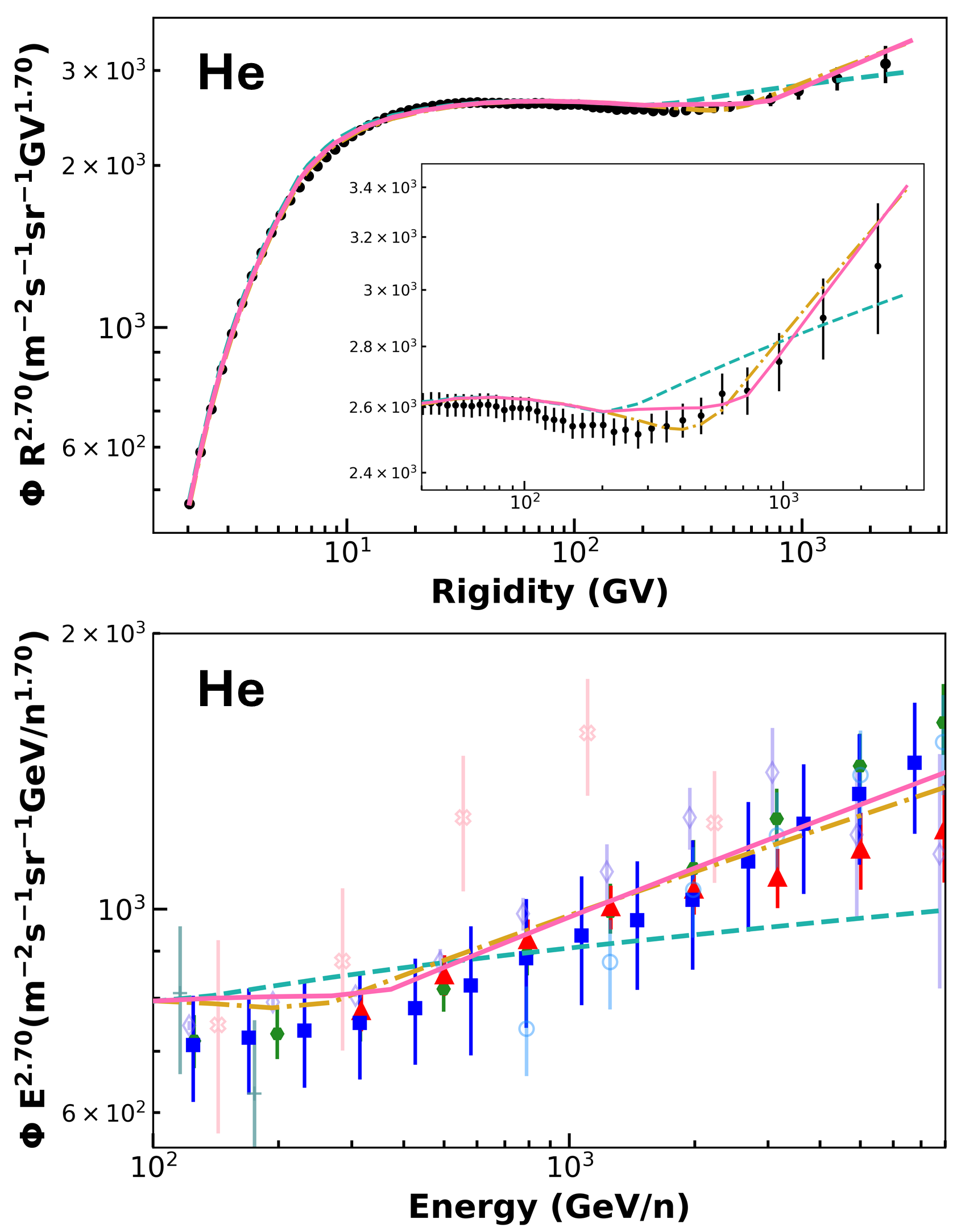}
    \caption{The results for the He data for Cases 1 (dashed teal line), 2 (dashed-dotted gold line), and 3 (solid pink line) are compared to the compiled data. A zoomed-in graph is shown above 40 GV in the top graph, also compared to the compiled data. 
    The results are modulated spectra with $\phi_{AMS}$ = $533 \pm 2$ MV.
    The legend can be found in Figure \ref{pHardeningFig}. 
    }
    \label{HeHardeningFig}

    \includegraphics[width=0.54\textwidth]{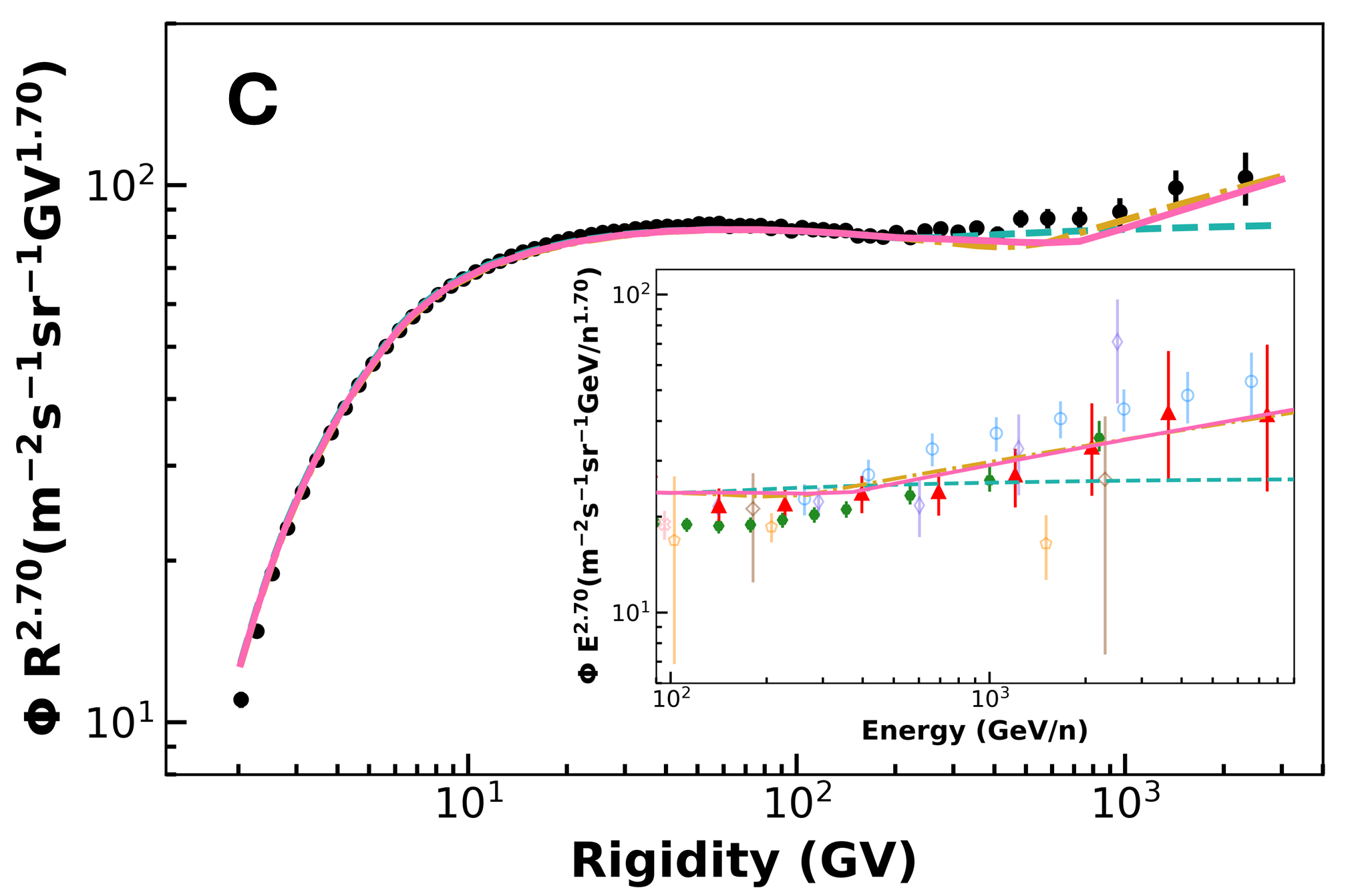}
    \caption{The results for the C data for Cases 1 (dashed teal line), 2 (dashed-dotted gold line), and 3 (solid pink line) are compared to the compiled data. An inserted graph is shown above 90 GeV/n with curves for Cases 1, 2, and 3, also compared to the compiled data.
    The results are modulated spectra with $\phi_{AMS}$ = $533 \pm 2$ MV.
    The legend can be found in Figure \ref{BCRatioFig}. The red upward-pointing filled triangles denote CREAM-II \citep{2009ApJ...707..593A} data in this figure.
    }
    \label{CFig}
\end{figure}

\begin{figure}[!h]
    \centering
    \includegraphics[width=0.56\textwidth]{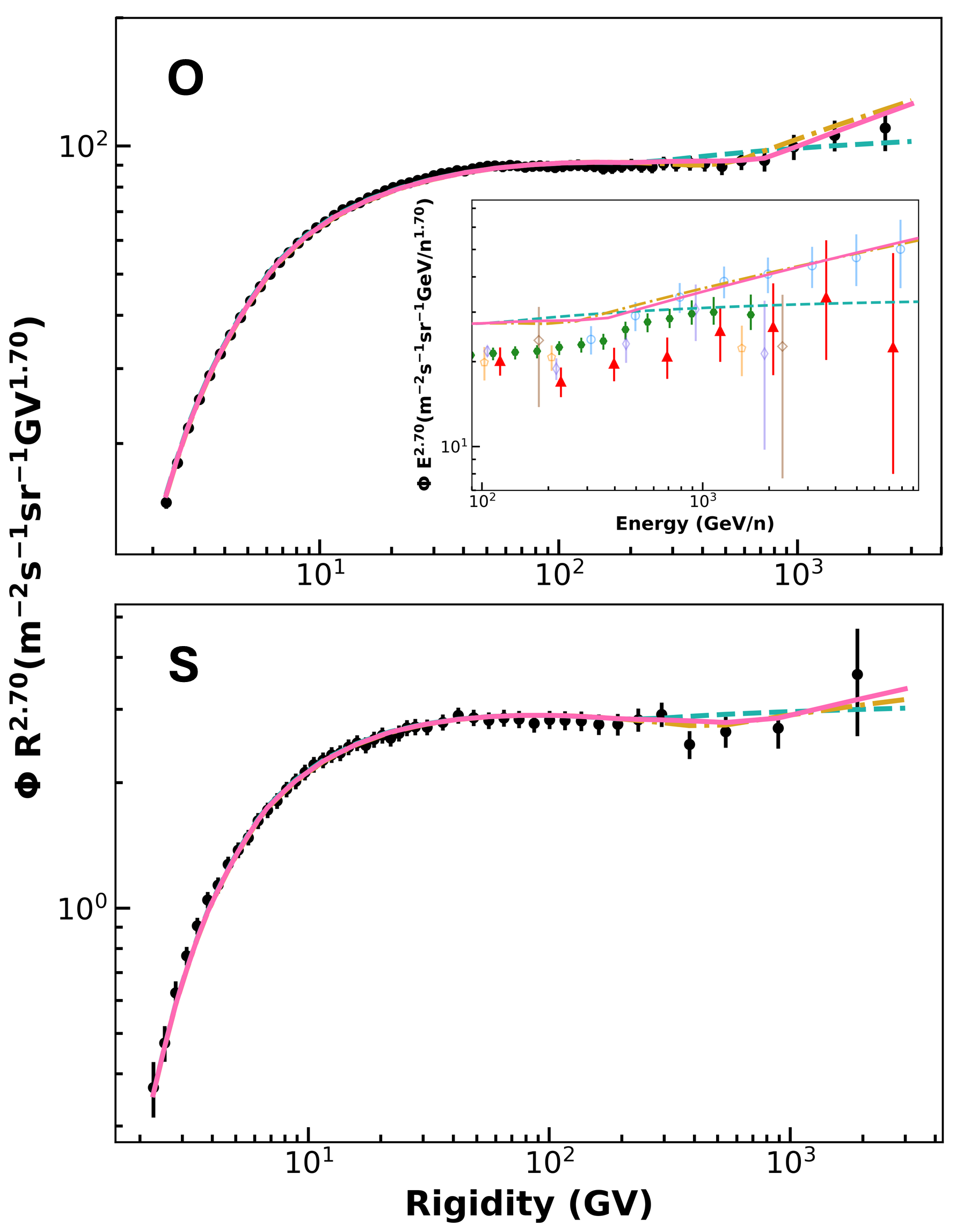}
    \caption{The results for the O and S data for Cases 1 (dashed teal line), 2 (dashed-dotted gold line), and 3 (solid pink line) are compared to the compiled data. The graph for O has an additional inserted graph shown above 90 GeV/n for Cases 1, 2, and 3, also compared to the compiled data. 
    The results are modulated spectra with $\phi_{AMS}$ = $533 \pm 2$ MV. 
    The legend can be found in Figure \ref{CFig}. 
    }
    \label{OandSFig}
\end{figure}

\begin{figure}[t]
    \centering
    \includegraphics[width=0.56\textwidth]{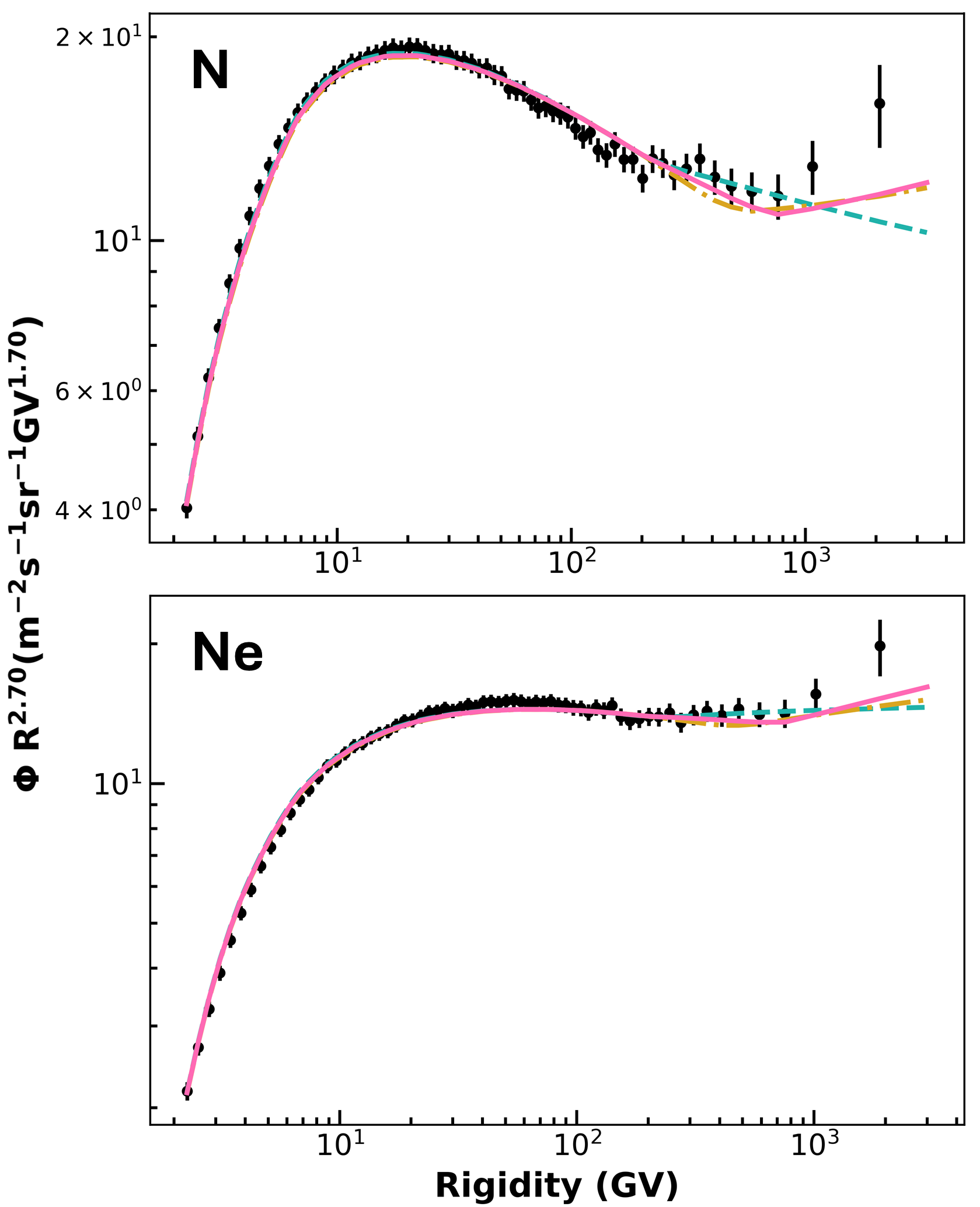}
    \caption{The results for the N and Ne data for Cases 1 (dashed teal line), 2 (dashed-dotted gold line), and 3 (solid pink line) are compared to the compiled data, the legend for which can be found in Figure \ref{BCRatioFig}. The results are modulated spectra with $\phi_{AMS}$ = $533 \pm 2$ MV.
    }
    \label{NandNeFig}
\end{figure}

\begin{figure}[!t]
    \centering
    \includegraphics[width=0.56\textwidth]{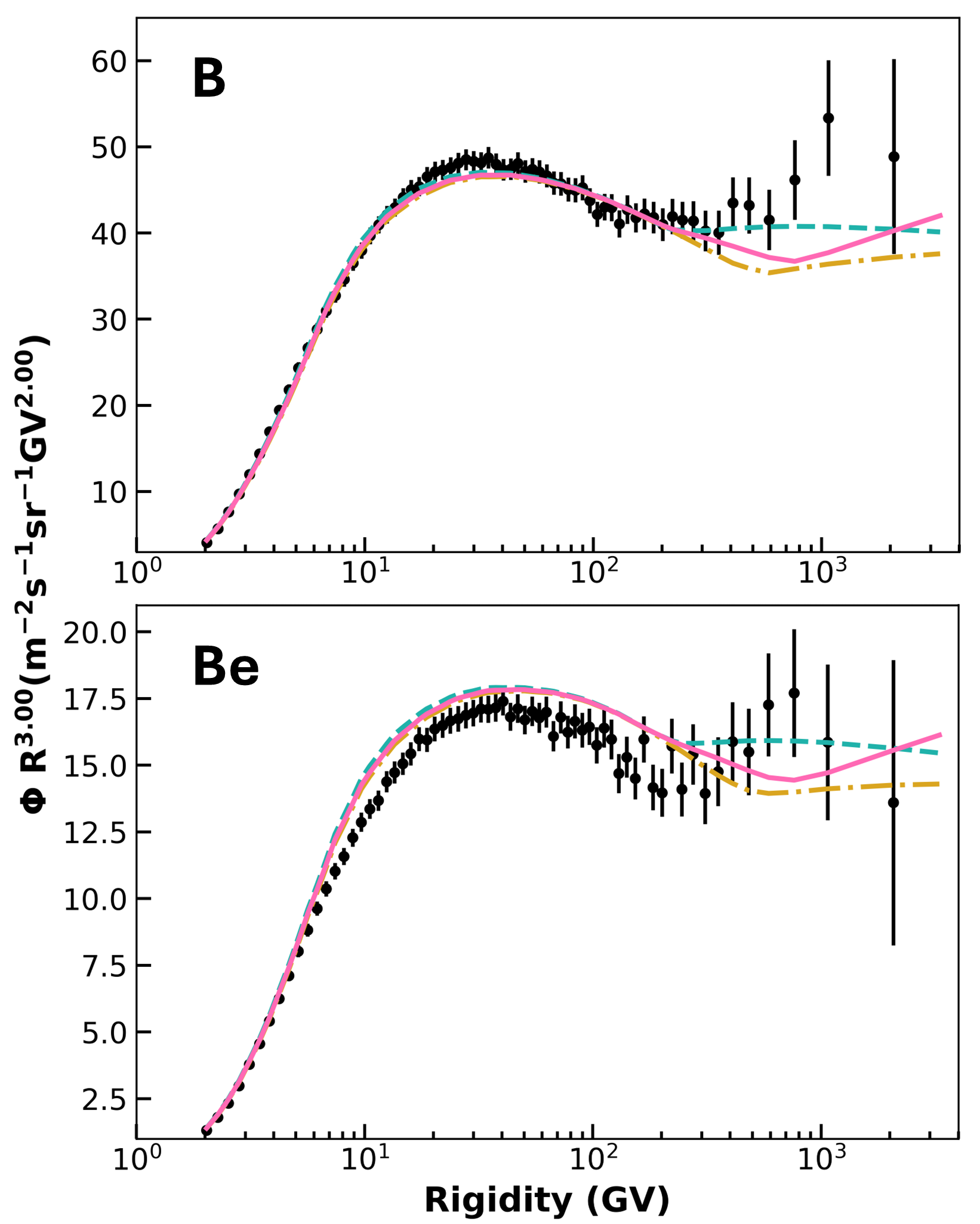}
    \caption{The results for the B and Be data for Cases 1 (dashed teal line), 2 (dashed-dotted gold line), and 3 (solid pink line) are compared to the compiled data, the legend for which can be found in Figure \ref{BCRatioFig}. The results are modulated spectra with $\phi_{AMS}$ = $533 \pm 2$ MV. 
    }
    \label{BandBeFig}
\end{figure}

\begin{figure}[t]
    \centering
    \includegraphics[width=0.56\textwidth]{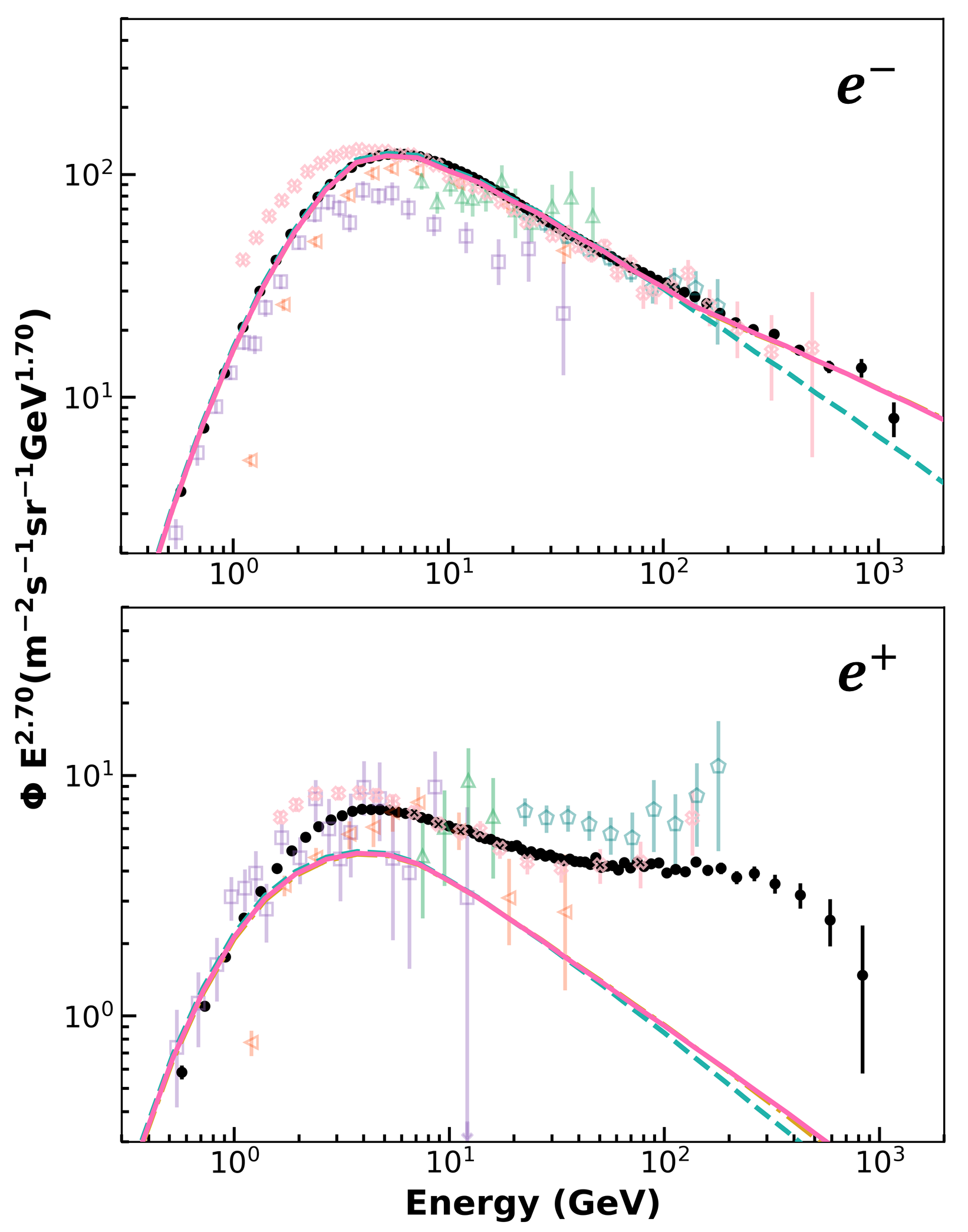}
    \caption{The results for the negative $e^{-}$ and $e^{+}$ data for Cases 1 (dashed teal line), 2 (dashed-dotted gold line), and 3 (solid pink line) are compared to the compiled data. 
    The results are modulated spectra with $\phi_{AMS}$ = $533 \pm 2$ MV. 
    The legend can be found in Figure \ref{BCRatioFig} with additional entries: purple open squares -- CAPRICE \citep{2000ApJ...532..653B}, teal open pentagons -- Fermi-LAT \citep{2012PhRvL.108a1103A}, orange right-pointing open triangles -- HEAT \citep{2001ApJ...559..296D}, green up-pointing triangles -- MASS \citep{2002A&A...392..287G}. 
    CAPRICE, Fermi-LAT, HEAT, MASS, and PAMELA data are shown for additional comparison along with the data used for optimization.
    }
    \label{ElecPosCasesFig}
\end{figure}

Results for the three cases for the B/C ratio, p, $\bar p$, He, C, O and S, N and Ne, B and Be, and negative $e^-$ and $e^+$ data are shown in Figures \ref{BCRatioFig}, \ref{pHardeningFig}, \ref{pbarFig}, \ref{HeHardeningFig}, \ref{CFig}, \ref{OandSFig}, \ref{NandNeFig}, \ref{BandBeFig}, and \ref{ElecPosCasesFig}, respectively. In all figures, the curve for Case 1 is denoted as the dashed teal line, the curve for Case 2 is denoted as the dash-dotted gold line, and the curve for Case 3 is denoted as the solid pink line. Since AMS-02 data are published in units of rigidity, the GALPROP-calculated curves are plotted in rigidity for comparison. Similarly, for other experiments that publish their data in kinetic energy per nucleon, another set of GALPROP-calculated curves are plotted in kinetic energy per nucleon. This separate plotting is done to avoid potential bias during unit conversion from rigidity to energy \citep[Appendix A]{2019A&A...627A.158D}. For experiments that publish their data in total energy, a conversion is done to kinetic energy per nucleon for comparison.

All three cases are found to be nearly identical below the diffusion coefficient break $\rho_1$ or injection spectra break $R_1$. The curves for the B/C ratio show a good fit to the data above $\sim$7 GV up to these breaks, as shown in Figure \ref{BCRatioFig}. However, the peak in the B/C ratio data at $\sim$4 GV is not as sharply reproduced by the curves. In the same rigidity range, p and C are slightly overproduced, as shown in Figures \ref{pHardeningFig} and \ref{CFig}, respectively, and B nuclei are slightly underproduced as shown in Figure \ref{BandBeFig}.

This smoother behavior in the B/C ratio data along with the behavior of p, B, and C at $\sim$4 GV is due to tensions from fitting these data simultaneously. The B/C ratio at $\sim$4 GV can be more accurately reproduced using only the B/C, B, and C data inputs. However, when p data are included in fitting, the B/C peak tends to be underproduced. This occurs because the optimizer gives preference to the p data due to its statistical significance.

In Figure \ref{BandBeFig}, the B nuclei are slightly underproduced from $\sim$20-40 GV. In the same figure, the Be nuclei are overproduced above $\sim$7 GV. 
This is most likely due to the total inelastic cross-section uncertainties of the B and Be isotopes \citep{2018PhRvC..98c4611G}, providing a constraint to the optimizer for the Be and B curves.

\clearpage 

The curves for the $\bar p$, He, O and S, N and Ne, and negative $e^-$ data show a good fit up to $\rho_1$ or $R_1$ in Figures \ref{pbarFig}, \ref{HeHardeningFig}, \ref{OandSFig}, \ref{NandNeFig}, and \ref{ElecPosCasesFig}, respectively. For $e^+$ in Figure \ref{ElecPosCasesFig}, the curves underproduce the data above $\sim$2 GV. The diffusion coefficient break at $\rho_{0} = 6.78 \pm 0.06$ GV is necessary to prevent an overproduction in the low energy $e^{+}$ data and produce a peak in the B/C ratio data at $\sim$4 GV \citep{Chen:2023eyy}.
A diffusion coefficient break below $\sim$ 10 GV at $3.586$
GV is used by \citet{PhysRevD.104.103029} and a break at $4.84-4.99$ GV is used by \citet{PhysRevD.108.063024}, which gives results that do not overproduce $e^{+}$ while producing the peak in the B/C ratio. 
Although not shown here, a good fit is also obtained for the Mg, Si, S, and Fe data up to $\rho_1$ or $R_1$ for all three cases.

All three cases reproduce the break and hardening in the B/C ratio data as seen in Figure \ref{BCRatioFig}. Despite this, not all three cases are able to fully reproduce the break and the hardening in the B and C data, as shown in Figures \ref{BandBeFig} and \ref{CFig} respectively. 
For p in Figure \ref{pHardeningFig}, Case 1 hardens less compared to the data above $\sim$600 GV. Cases 2 and 3 both reproduce the break and the spectral hardening in the p data. However, all three cases do not produce enough excess as seen in the $\bar p$ data above $\sim$100 GeV in Figure \ref{pbarFig}. This implies that the excess in the $\bar p$ data cannot be explained by a diffusion coefficient break, injection spectra breaks, or a combination of both. Note that previous studies, including \citet{PhysRevResearch.2.023022} and \citet{PhysRevD.104.103029}, have produced an enhanced $\bar p$ fit at this energy.  \citet{PhysRevResearch.2.023022} used USINE v3.5 and incorporated uncertainties from production cross-sections, transport in the galaxy, and the fit to parent species fluxes. \citet{PhysRevD.104.103029} used GALPROP v56, considered acceleration of secondary antiprotons in old supernova remnants, and considered cross-section parameterizations based on collider data for secondary $\bar p$ production. In this work, GALPROP v57 is used with the cross-section parametrizations detailed in \citet[][\& references therein]{Porter_2022} and cross-section option “Opt 022”, which utilizes phenomenological approximations by \citet{1998ApJ...501..911S}.

Case 1 produces a lower rigidity break compared to the He data in Figure \ref{HeHardeningFig}. Case 1 also does not produce enough hardening compared to to the He data. For C in Figure \ref{CFig}, Case 1 reproduces the break but hardens less than this element above $\sim$800 GV. Case 1 can reproduce the break and spectral hardening seen in the O and S data, as seen in Figure \ref{OandSFig}. However, it also produces a flatter curve compared to these two elements. In contrast to Case 1, Cases 2 and 3 both reproduce the break and the spectral hardening in the He, C, O, and S data.

For N in Figure \ref{NandNeFig}, Case 1 hardens less than the data above $\sim$800 GV. For Ne in the same figure, Case 1 hardens less than the data above $\sim$1 TV. Cases 2 and 3 reproduce the break but harden less than these two elements above $\sim$800 GV, as shown in Figure \ref{NandNeFig}. 

\enlargethispage{\baselineskip}

Case 1 produces the break seen in the Be data, as seen in Figure \ref{BandBeFig}. However, Case 1 produces a curve that has a slightly flatter slope compared to the Be data above the break. In the same figure, Case 1 can also produce the break seen in the B data. However, Case 1 produces a curve that has a flatter slope compared to the B data above the break. Case 2 produces the break seen in the Be data, and it is able to produce a curve with a slope similar to the Be data above the break. However, Case 2 produces a higher rigidity break compared to the B data. It also produces a curve with a slightly flatter slope compared to the B data above the break. Case 3 produces a higher rigidity break compared to the Be data. However, it is able to produce a curve with a slope that is comparable to the Be data above the break. Case 3 produces a higher rigidity break compared to the B data as well. However, it is able to produce a curve with a slope that is comparable to the B data above the break. 

The overproduction in the Be nuclei above $\sim$7 GV potentially implies that Cases 2 and 3 need their break at a lower rigidity. This overproduction also potentially implies that Case 1 produces less hardening compared to the Be data.

In Figure \ref{ElecPosCasesFig}, Case 1 does not produce enough excess for the negative $e^{-}$ data above $\sim$100 GeV. However, Cases 2 and 3 are able to produce enough excess for the $e^{-}$ data above $\sim$100 GeV. In contrast, all three cases underproduce the $e^{+}$ data above $\sim$2 GeV in Figure \ref{ElecPosCasesFig}. This implies that the excess seen in $e^+$ data cannot be explained by a diffusion coefficient break, injection spectra breaks, or a combination of both.

Though not shown here, all three cases both reproduce the break and the spectral hardening in the Mg and S data. Additionally, all three cases both produce a break and spectral hardening that is consistent with the Si and Fe data.

\subsection{Proton and Helium Spectra} \label{softeningSubsection} 

\begin{figure}[h]
    \centering
    \includegraphics[width=0.56\textwidth]{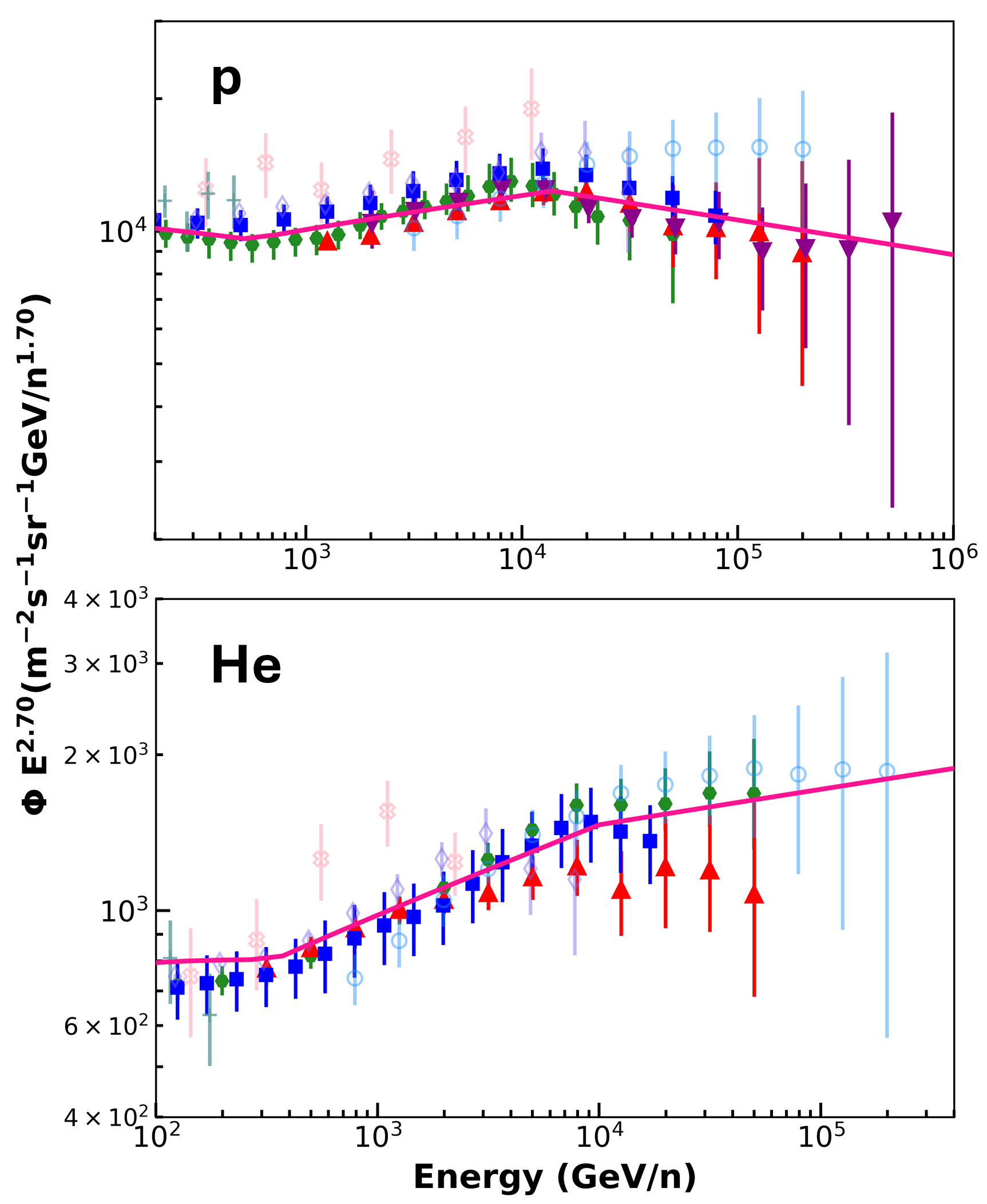}
    \caption{The results for the p and He data, after fitting above $R_2$ (solid pink line), are compared with the compiled data, the legend for which can be found in Figure \ref{pHardeningFig}. 
    }
    \label{SofteningFig}
\end{figure}

The best-fit parameters and their associated error for fitting the p and He data above $R_2$ are obtained. These results are the outcome of Stage 3 from Section \ref{methoPandHeSection}. 
Results of this p and He fit are built off of Case 3.  
The injection spectrum for p has a second injection spectral break $R_2$ at $13.89 \pm 0.02$ TV with an injection spectral index of $\gamma_{3}$ = $2.334 \pm 0.029$ above the break. The injection spectrum for He has a break $R_{2}$ at $20 \pm 1$ TV and an index $\gamma_{3}$ of $2.188 \pm 0.049$. 
Using the injection spectra results of Case 3, the index $\gamma_{2}$ for He is smaller by $\sim$0.1 compared to p from 687 GV to 13.89 TV: He is harder than p in this rigidity range. Results for the p and He data are shown in Figure \ref{SofteningFig}. The break $R_2$ and index $\gamma_3$ for p and He are consistent with these data.

No smoothing parameters were used for the injection spectra power law for the p and He fit, which serves as a potential limitation. 
Reasons behind the softening in the p and He data could be due to -- but are not limited to -- additional sources, as shown by \citet{gaisser2013cosmic} and \citet{BOWMAN20222703}, or a local reacceleration of particles out of background cosmic rays by a nearby shock, as shown by \citet{malkov2022origin}.

\subsection{Positron Spectrum} \label{posExcessSubsection}

The best-fit parameters and their associated error for fitting the positron data with a charge-symmetric primary positron source are obtained. These results are the outcome of Stage 4 from Section \ref{methoPosSection}. Results of this charge-symmetric primary $e^{+}$ source are built off of Case 3.

Table \ref{chargeSymmPrimarySourceInject} summarizes the injection spectral break rigidity $R_{0}$, the spectral index $\gamma_{0}$ below $R_{0}$, the break $R_1$, the index $\gamma_{1}$ between $R_{0}$ and $R_1$, and the index $\gamma_2$ above $R_1$ obtained for the charge-symmetric primary $e^{+}$ source and the primary e$^{-}$ source. The abundance for the primary $e^{+}$ source is $97.2 \pm 0.9$ relative to the negative $e^{-}$ abundance at $1.06 \cdot 10^{3}$.

\begin{figure}[!h]
    \centering
    \includegraphics[width=0.56\textwidth]{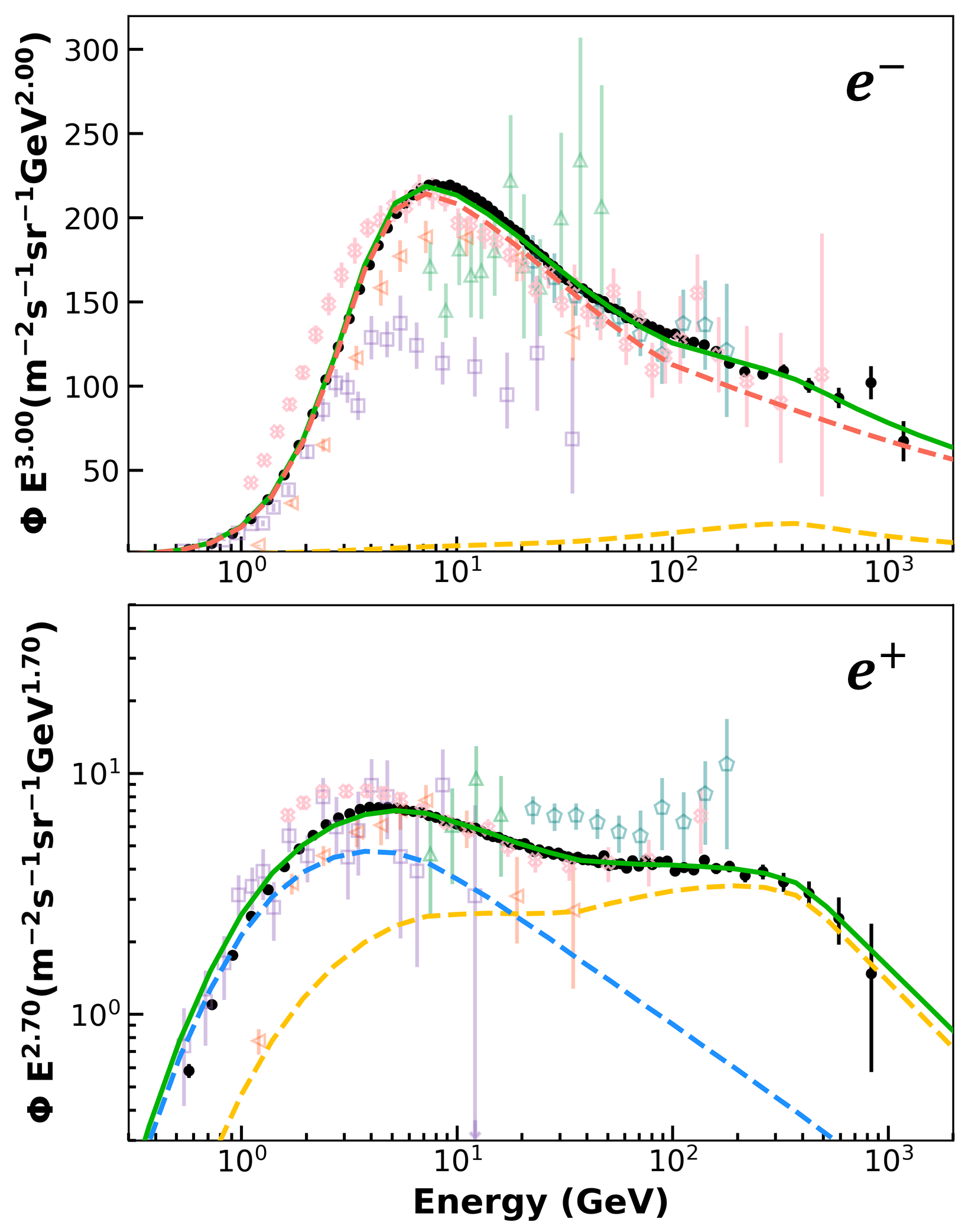}
    \caption{The results for the negative $e^{-}$ and $e^{+}$ data after introducing a charge-symmetric primary source are compared to the compiled data, the legend for which can be found in Figure \ref{ElecPosCasesFig}. In the top graph, the primary $e^{-}$ result (dashed red line), charge-symmetric $e^{+}$ result (dashed yellow line), and sum of these two results (solid green line) are shown. In the bottom graph, the secondary $e^{+}$ result (dashed blue line), charge-symmetric $e^{+}$ result (dashed yellow line), and sum of these two results (solid green line) are shown. The results are modulated spectra with $\phi_{AMS}$ = $533 \pm 2$ MV. 
    }
    \label{posExcessPrimaryFig}
\end{figure}

Figure \ref{posExcessPrimaryFig} shows the results for this charge-symmetric primary $e^{+}$ source for the negative $e^{-}$ and $e^{+}$ data. An introduction of a charge-symmetric primary source is able to describe the positron data. This source is also compatible with the negative $e^{-}$ data.

\subsection{All-Particle Spectrum} \label{APSpectrumSubsection}

\begin{figure}[h]
    \centering
    \includegraphics[width=0.625\textwidth]{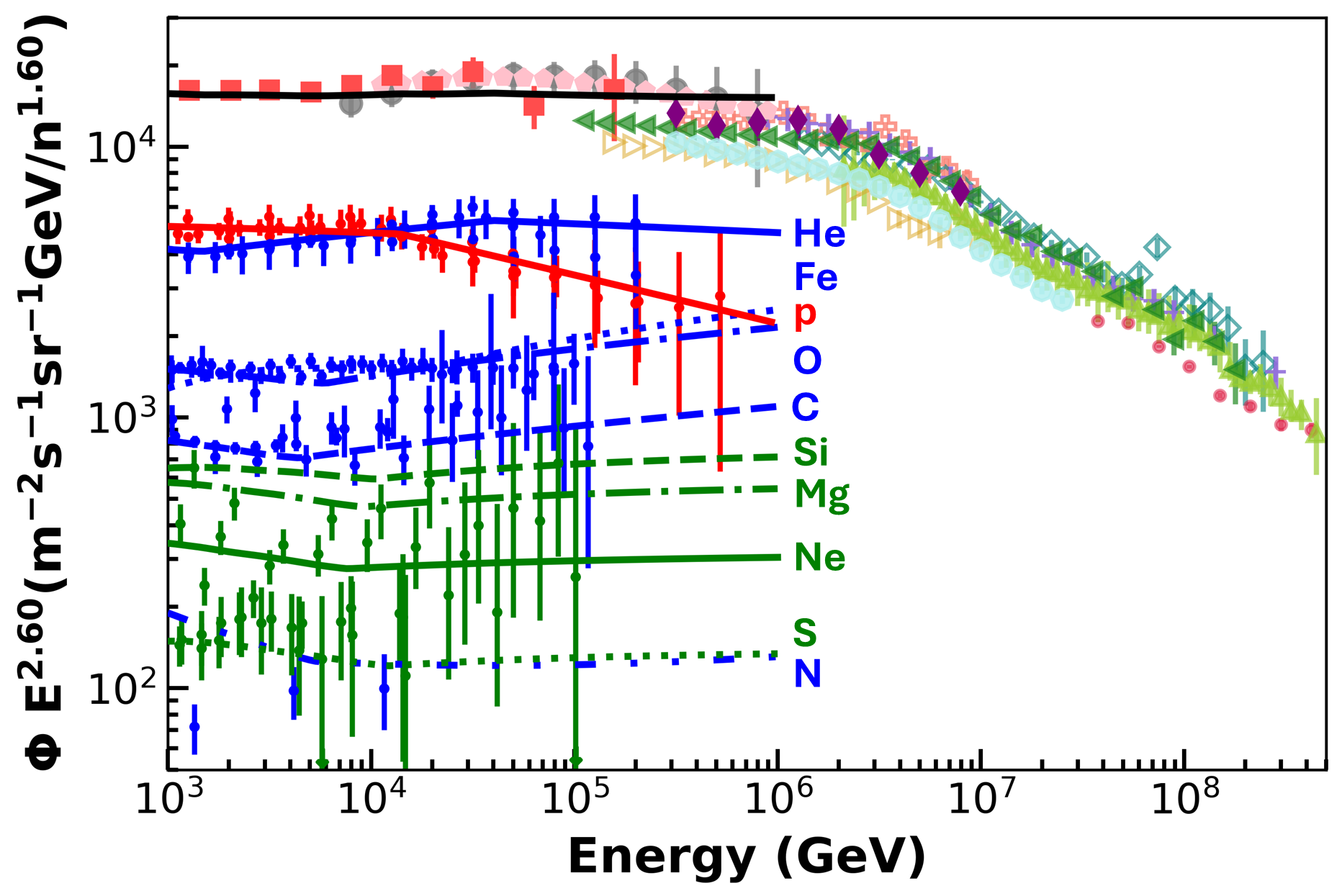}
    \caption{The calculated all-particle spectrum (black solid line) is compared with the compiled data. GALPROP v57 results for p (solid red line), He (solid blue line), C (dashed green line), N (dashed-double-dotted blue line), O (dash-dotted blue line), Ne (solid green line), Mg (dashed-dotted green line), Si (dashed green line), S (dotted green line), and Fe (dotted blue line) are also compared with the compiled data. 
    All-particle data legend: red filled squares -- ATIC-2 \citep{Panov_2009}, gold right-pointing open triangles -- CASA-MIA \citep{GLASMACHER1999291}, teal open wide diamonds -- GAMMA \citep{2014PhRvD..89l3003T}, 
    light pink filled pentagons -- HAWC QGSJET-II-04 \citep{2022icrc.confE.330M}, orange open thick pluses -- HEGRA \citep{2000A&A...359..682H}, purple filled diamonds -- IceTop QGSJET-II-04 \citep{2020PhRvD.102l2001A}, 
    burgundy filled dots -- KASCADE-GRANDE \citep{2017ICRC...35..316A}, 
    light blue filled octagons -- LHAASO Sibyll 2.3d \citep{PhysRevLett.132.131002}, grey filled circles -- NUCLEON-KLEM \citep{2019AdSpR..64.2546G}, yellow green upward-pointing filled triangles -- TALE \citep{2018ApJ...865...74A}, medium green right-pointing filled triangles -- TIBET QGSJET-I+HD \citep{Amenomori_2008}, 
    and light purple thin pluses -- TUNKA-133 \citep{2014NIMPA.756...94P}. 
    Direct measurement data are shown as solid dots: p, He, C, O, and Fe from CALET \citep{PhysRevLett.125.251102, PhysRevLett.126.241101, PhysRevLett.129.101102, PhysRevLett.130.171002}; p and He from CREAM-I+III \citep{2017ApJ...839....5Y}; C, N, O, Ne, Mg, and Si from CREAM-II \citep{2009ApJ...707..593A}; p and He from DAMPE \citep{2019SciA....5.3793A}; p from ISS-CREAM \citep{2022ApJ...940..107C}; S from \citet{1997APh.....6..155K} and \citet{1993PhRvD..48.1949I}.}
    \label{allPartFig}
\end{figure}

After obtaining the results of Sections \ref{Stage2MethodSect}, \ref{methoPandHeSection}, and \ref{methoPosSection}, the all-particle spectrum is calculated up to GALPROP v57's limit of $\sim$1 PeV \citep{Porter_2022}. Only the hardening from Case 3 is considered for the calculation due to its use for the results of Sections \ref{softeningSubsection} and \ref{posExcessSubsection}. To calculate the all-particle spectrum from GALPROP v57, the same total energy bins are first obtained across each cosmic ray species. The flux from GALPROP v57 for each cosmic ray species is then interpolated. The sum of these interpolated spectra results in the all-particle spectrum, shown as the solid black line in Figure \ref{allPartFig}. The results for Cases 1, 2, and 3 and the results from Sections \ref{softeningSubsection} and \ref{posExcessSubsection} are compatible with the all-particle spectrum.

For the elements of C, N, O, Ne, Mg, Si, S, and Fe, the current experimental data show only hardening due to their limited energy reach. 
Potential softening at higher energies for these elements,
as seen in the current p and He data, has not yet been observed nor calculated in this work. Consequently, above $\sim$200 TeV, the calculated all-particle spectrum can be considered as an upper limit. Comparing the individual elemental contributions to the all-particle spectrum at $\sim$700 TeV, p contributes $\sim$16\%, He contributes $\sim$32\%, and C + O + Fe contributes $\sim$37\%. This suggests a potential overproduction of the all-particle spectrum due to elements $Z>2$ approaching 1 PeV. 

With both the diffusion coefficient and injection spectra breaks for the spectral hardening being rigidity-dependent, the elements are seen to have different breaks in total energy. This difference in total energy breaks has resulted in the breaks compensating with each other once summed together, leading to the overall shape of the all-particle spectrum being flat.

If the softening break in p occurs at the same rigidity across the elements of C through Fe, the corresponding break would be at $\sim$84 TeV for C, $\sim$112 TeV for O, and $\sim$214 TeV for Fe. 
This elemental softening is expected to produce a decrease in the all-particle data above $\sim$1 PeV, which may better align with the experimental data.

\enlargethispage{\baselineskip}

\pagebreak

\begin{table}[h]
    \centering
    \caption{Injection spectra results for the negative $e^{-}$ primary source and charge-symmetric primary $e^{+}$ source.}
    \begin{tabular}{ccc} \dtoprule
         & $e^{-}$ Primary & $e^{\pm}$ Charge-Symmetric\\ \midrule
         $\gamma_{0}$ & $1.492 \pm 0.026$ & $2.258 \pm 0.010$ \\
         $R_{0}$ (GV) & $4.48 \pm 0.06$ & $38 \pm 4$ \\
         $\gamma_{1}$ & $2.703 \pm 0.006$ & $1.801 \pm 0.038$ \\
         $R_{1}$ (GV) & $99 \pm 3$ & $412 \pm 178$ \\
         $\gamma_{2}$ & $2.504 \pm 0.012$ & $2.862 \pm 0.029$ \\ \dbottomrule
    \end{tabular}
    \label{chargeSymmPrimarySourceInject}
\end{table}

\section{Discussion} \label{discussSect}

It is noted that a smaller value of the index $\delta_2$ in Case 1 with the same diffusion coefficient break $\rho_1$ could potentially produce a slope above the break that is more consistent with the data. However, this would produce tension with fitting the break $\rho_1$ itself. Adjusting the break $\rho_1$ to counteract this may still be unable to produce a break at an appropriate rigidity across all species. Introducing a smoothing parameter to attempt to address this could be included in future work. Additionally, it is noted that the value of the break $\rho_1$ and the index above the break $\delta_2$ is mainly constrained by the observational data from AMS-02 due to its statistical significance.

These results for Cases 1, 2, and 3 are an update to our previous work in \citet{Wu:2021ldb}, where the same cases were also explored. Results in \citet{Wu:2021ldb} were obtained in GALPROP v56 by hand-tuning, while results in this work are obtained using GALPROP v57's parameter optimization module. More recent data from the same experiments used by \citet{Wu:2021ldb}, additional He data from CALET, and additional S data from AMS-02 are used in this work. Only one diffusion coefficient break was considered in \citet{Wu:2021ldb}, while two are considered in this work. The second injection spectra breaks $R_1$ differ between our studies: in this work, $R_1$ for p is found to be $\sim$160-250 GV lower, and $R_1$ for elements $Z \geq 2$ are found to be $\sim$160-450 GV higher. 

This work also serves as an update to our previous work in \citet{Chen:2023eyy}, where the same cases were also explored. In our previous work, results were obtained by hand-tuning in GALPROP v57 for elements p through O. Additional He data from CALET is included in this work compared to our previous work.

A study of using only injection spectra breaks to fit the spectral hardening was performed in \citet{Johannesson_2016}, analogous to Case 2 in this work. For their study, a Bayesian search using the MultiNest sampling algorithm and the BAMBI neural network machine learning package was done using GALPROP v54 and with ACE-CRIS, CREAM-I, CREAM-II, HEAO3-C2, and PAMELA data. In this work, MINUIT is used with GALPROP v57 with AMS-02, CALET, CREAM-I, CREAM-I+III, CREAM-II, DAMPE, ISS-CREAM, and NUCLEON-KLEM data. Two sets of results are included in their work: one set for elements p, $\bar p$, and He, and another set for elements Be through Si. In contrast, this work has elements Z $\geq$ 1 in three different groups for injection spectra results, as summarized in Table \ref{hardeningSourceInjection}. Separate diffusion coefficient results were derived for the two sets of elements in their work, while all species have the same diffusion coefficient results below $\rho_1$ in this work. The injection spectrum break for the spectral hardening is set at 220 GV in their work, while injection spectra breaks are allowed to vary in this work. All injection spectra breaks $R_1$ are $\sim$100-200 GV higher in this work for Case 2 in comparison to their work for either set of results. While the injection spectral index $\gamma_2$ above the break $R_1$ is similar for p, Ne, Mg, Si, and S, the index $\gamma_2$ above $R_1$ is smaller by $\sim$0.1 for He, C, N, O, and Fe in this work.

A study of having only a diffusion coefficient break and only injection spectra breaks for fitting the spectral hardening was done in \citet{Boschini_2020}, referred to as the "P-Scenario" and the “I-Scenario” in the paper, respectively. In comparison to this work, an MCMC method was used with GALPROP v56, instead of MINUIT with GALPROP v57, using earlier AMS-02 data. The high-rigidity diffusion coefficient break $\rho_1$ from Case 1 in this work is $\sim$170 GV lower in comparison to the one in the P-Scenario, and the diffusion coefficient index above the break in Case 1 $\delta_2$ is $\sim$0.23 larger in comparison to the index from the P-Scenario. 

Fitting the p and He data above the injection spectra break $R_2$ in this work serves as an update to our previous work in \citet{Chen:2023eyy}, where the same injection spectra parameters were introduced to fit the p and He data. The same differences between these works as described earlier in this section also hold here. 
The injection spectrum break $R_2$ for He in this work is $\sim$10 TV higher than what is in our previous work.

A study of using an injection spectrum break for the p and He data in GALPROP was also done by \citet{2020ApJS..250...27B}. In comparison to this work, an MCMC method was used with GALPROP v56 using earlier AMS-02 data. 
The injection spectrum break $R_2$ for He in this work is $\sim$10 TV lower, and the injection spectrum index above the break $\gamma_3$ for He is $\sim$0.18 smaller.

The fitting of the positron data in this work serves as an update to our previous work in \citet{Chen:2023eyy}, where a primary positron source was also introduced. In this work, the primary positron source is assumed to be charge-symmetric, while in our previous work it is not. The same differences between these works as described earlier in this section also hold here. For the primary $e^+$ source injection spectrum, two breaks are obtained in this work in comparison to one break in our previous work.

Previous studies, such as \citet{DELLATORRE201527} using GALPROP v54 and \citet{De_La_Torre_Luque_2023} using DRAGON2 with FLUKA cross-sections, used a primary pulsar source to describe the positron excess. In this work, GALPROP v57 is used instead with a primary supernova remnant source. Both of these studies used a single break in the power law for the source, while two breaks are used in the power law in this work. Future work exploring a pulsar source instead of a supernova remnant is considered.

\section{Conclusion} \label{concluSect}

Using GALPROP v57's parameter optimization routine, best-fit diffusion coefficient and injection spectra results are obtained for recent cosmic ray data. 

Having only a high-rigidity diffusion coefficient break creates a lower rigidity break in comparison to the p and He data, and it also does not produce enough hardening for these two elements. Having only a high-rigidity diffusion break also does not produce enough hardening for the C, N, and Ne data. It also produces a flatter result compared to the O and S data. It is able to produce the break seen in the Be and B data, but it creates results that have a flatter slope in comparison to these two elements above their respective break. Additionally, the excess in the negative $e^-$ data above $\sim$100 GeV is not produced by only having a high-rigidity diffusion coefficient break. 

Having only injection spectra breaks and a combination of both a diffusion coefficient break and injection spectra breaks reproduces a break and the spectral hardening for the p, He, C, O, and S data. Both of these cases produce less hardening compared to the N and Ne data above $\sim$700 GV. Both of these cases also produce a break at a higher rigidity compared to the Be and B data. However, a combination of a diffusion coefficient break and injection spectra breaks produces a slope that is comparable to the Be and B data above the break. Both of these cases also produce the excess seen above $\sim$100 GeV in the negative $e^-$ data. 

\pagebreak

Having a diffusion coefficient break, injection spectra breaks, or a combination of both are unable to reproduce the $\bar p$ data above $\sim$100 GeV and the $e^+$ data above $\sim$2 GeV. The inclusion of a charge-symmetric primary $e^{+}$ source is able better to describe the positron data and can be successfully incorporated into the negative $e^{-}$ spectrum. 

The introduction of an additional higher-energy injection spectrum break for the p and He data returns results that are consistent with the data. 

An upper limit to the all-particle spectrum is estimated from $\sim$200 TeV to $\sim$1 PeV. Though there is a softening for the p and He spectra, the spectra for C through Fe exhibit only hardening due to their limited energy range. As a result, the all-particle spectrum is flat approaching 1 PeV. An elemental softening for C through Fe is expected to produce a decrease in the all-particle spectrum above $\sim$1 PeV. 

Future work aims to improve the fit to the excess seen above $\sim$100 GeV in the $\bar p$ data. A pulsar source to describe the positron excess will also be investigated. Future work also includes incorporating a more dedicated treatment of solar modulation using HelMod \citep{BOSCHINI20182859}. Further statistical analysis to explore whether the data favor a high-rigidity diffusion coefficient break, injection spectra breaks, or a combination of both breaks is also considered for future development.


\small
\printbibliography 

\end{document}